\documentclass{iopart}
\usepackage{graphicx}
\usepackage{iopams}
\usepackage[square,numbers,sort&compress]{natbib}
\usepackage{aas_macros}
\providecommand{\newblock}  

\begin{document}

\title[Anisotropies from dark matter substructure]{Revealing dark matter substructure with anisotropies in the diffuse gamma-ray background}

\author{Jennifer M. Siegal-Gaskins\footnote{Present address: Center for Cosmology and Astro-Particle Physics, The Ohio State University,
191 W.~Woodruff Avenue, Columbus, OH 43210; \tt{jsg@mps.ohio-state.edu}}}
\address{Kavli Institute for Cosmological Physics and Department of Physics,\\ University of Chicago, 5640 S.~Ellis Avenue, Chicago, IL 60637}
\ead{jsg@kicp.uchicago.edu}

\begin{abstract}

The majority of gamma-ray emission from Galactic dark matter annihilation is likely to be detected as a contribution to the diffuse gamma-ray background.  I show that dark matter substructure in the halo of the Galaxy induces characteristic anisotropies in the diffuse background that could be used to determine the small-scale dark matter distribution.  I calculate the angular power spectrum of the emission from dark matter substructure for several models of the subhalo population, and show that features in the power spectrum can be used to infer the presence of substructure.  The shape of the power spectrum is largely unaffected by the subhalo radial distribution and mass function, and for many scenarios I find that a measurement of the angular power spectrum by Fermi will be able to constrain the abundance of substructure.  An anti-biased subhalo radial distribution is shown to produce emission that differs significantly in intensity and large-scale angular dependence from that of a subhalo distribution which traces the smooth dark matter halo, potentially impacting the detectability of the dark matter signal for a variety of targets and methods.

\end{abstract}


\section{Introduction}
\label{sec:intro}

A wealth of evidence suggests that most of the matter in the universe is in the form of non-baryonic dark matter \citep[see][for reviews]{bergstrom_00, bertone_hooper_silk_05}, but the fundamental nature of this constituent remains unknown.
Theoretical work has produced several candidates, of which many current favorites fall under the category of weakly interacting massive particles (WIMPs).  These include the lightest supersymmetric particle (LSP, often termed the neutralino) in supersymmetric extensions to the standard model, and the lightest Kaluza-Klein particle (LKP) arising in theories of universal extra dimensions.  Both the neutralino and the LKP are stable, but can produce photons and other standard model particles through self-annihilation.  Dark matter could thus be detected indirectly by observing these annihilation products.  

The prospects for detecting dark matter via gamma-rays from annihilation have been studied extensively.  The Galactic Center \cite{berezinsky_bottino_mignola_94, dodelson_hooper_serpico_08, serpico_zaharijas_08}, dwarf galaxies \cite{baltz_briot_salati_etal_00, strigari_koushiappas_bullock_etal_08}, dark matter subhalos \cite{baltz_taylor_wai_07,  pieri_bertone_branchini_08, kuhlen_diemand_madau_08}, and intermediate mass black holes \cite{bertone_zentner_silk_05}, along with Galactic \cite{bergstrom_edsjo_gondolo_etal_99, calcaneo-roldan_moore_00, berezinsky_dokuchaev_eroshenko_03} and extragalactic \cite{bergstrom_edsjo_ullio_01, ullio_bergstrom_edsjo_etal_02, elsasser_mannheim_05, oda_totani_nagashima_05} diffuse emission, have all been considered as targets for indirect searches.  
However, detection methods which rely in whole or in part on the amplitude of the signal are hindered by uncertainties in the properties of the dark matter particle and by limited knowledge of the distribution of dark matter, particularly its clustering properties on small scales.  Even methods which make use of spectral information to extract the signal must contend with the presence of uncertain astrophysical foregrounds.

Recent work has proposed that angular features in the diffuse extragalactic gamma-ray background (EGRB) could be used to identify emission from dark matter.  An initial calculation of the angular power spectrum of the EGRB from dark matter annihilation in large-scale structures was made by \citet{ando_komatsu_06}, and subsequent studies have further investigated this approach \cite{ando_komatsu_narumoto_etal_07a, cuoco_hannestad_haugbolle_etal_07, cuoco_brandbyge_hannestad_etal_07}.  Estimates of the angular power spectrum of the EGRB from source classes other than dark matter have also been derived \cite{zhang_beacom_04, miniati_koushiappas_di-matteo_07, ando_komatsu_narumoto_etal_07b}, providing a basis for comparison.  
While the large-scale distribution of dark matter is fairly well understood, the structure of dark matter halos on subgalactic scales, which is the focus of this study, is only minimally constrained by current observations.  The intrinsic properties of the dark matter particle influence the formation and survival of substructure within galaxy-sized halos, and thus the small-scale distribution of dark matter remains an important test of its fundamental nature.  

For a generic WIMP cold dark matter (CDM) candidate, free-streaming and collisional damping induce a lower bound on the minimum mass of a CDM structure at around $10^{-6}$ to $10^{-4}$ M$_{\odot}$ \cite{green_hofmann_schwarz_05,loeb_zaldarriaga_05, bertschinger_06}, suggesting that the Galactic halo may be populated by an enormous number of dark matter clumps of roughly Earth-mass.  However, it is unclear whether the smallest halos are resilient enough to survive until the present day.  This issue has been investigated in numerical simulations  \cite{diemand_moore_stadel_05, diemand_kuhlen_madau_06} and analytically \cite{berezinsky_dokuchaev_eroshenko_03, berezinsky_dokuchaev_eroshenko_06, zhao_hooper_angus_etal_07, berezinsky_dokuchaev_eroshenko_08}, but has not yet been tested observationally.  In addition, numerical simulations predict far more subhalos in a galaxy-sized halo than the number of known luminous satellites of the Milky Way, spawning several proposed mechanisms for reducing small-scale structure.  For example, the abundance of substructure may be decreased in the CDM framework by a suppression of small-scale power in the primordial power spectrum \cite{kamionkowski_liddle_00, zentner_bullock_03}.

Modifications of the properties of the dark matter particle can also alter predictions for structure on subgalactic scales.  Self-interacting dark matter (SIDM) \cite{spergel_steinhardt_00, yoshida_springel_white_etal_00, dave_spergel_steinhardt_etal_01, colin_avila-reese_valenzuela_etal_02},
proposed to bring predictions for halo structure better in line with observations, relaxes the assumption that dark matter is collisionless by introducing self-interaction to standard CDM candidates via a non-negligible scattering cross-section.  While inheriting the successes of standard CDM on large scales, SIDM leads to less substructure in galaxy-sized halos, and halos with shallower inner density profiles.  Since SIDM candidates are variations of standard CDM candidates, they could similarly be detected by gamma-ray emission from annihilation.  
Warm dark matter (WDM) \cite{hogan_dalcanton_00, colin_avila-reese_valenzuela_00, bode_ostriker_turok_01} and MeV-scale dark matter \cite{hooper_kaplinghat_strigari_etal_07} have also been invoked to reduce structure on small-scales.  In these scenarios candidates are typically too light to produce gamma-ray emission, but may be detectable at lower energies. 

The work presented here investigates the prospects for using anisotropies in the diffuse gamma-ray background to gain new insights into the Galactic dark matter distribution.  In contrast with previous studies of large-scale angular features from Galactic dark matter \cite{dixon_hartmann_kolaczyk_etal_98, hooper_serpico_07, berezinsky_dokuchaev_eroshenko_07}, this approach focuses on fluctuations at small angular scales as an observational probe of Galactic dark matter substructure. 
Previous work has predicted that most subhalos will not be detectable individually by the Fermi Gamma-ray Space Telescope; this study suggests how Fermi could overcome this difficulty by using angular correlations in diffuse emission to statistically infer the presence of a subhalo population.

In many scenarios the annihilation signal from Galactic dark matter is expected to dominate over that from extragalactic dark matter \cite{hooper_serpico_07}, and although emission from a Galactic dark matter component has by no means been indisputably found, possible indications have already been identified: \citet{dixon_hartmann_kolaczyk_etal_98} presented evidence of a halo feature in the EGRET data, and \citet{hooper_finkbeiner_dobler_07} proposed that the observed excess microwave emission in the inner Galaxy \cite{finkbeiner_04, dobler_finkbeiner_08}, dubbed `the WMAP haze', is in fact synchrotron emission from electrons and positrons produced by dark matter annihilation.
Moreover, even if the Galactic dark matter contribution at high latitudes is modest, it is necessarily a foreground to any measurement of the EGRB, and so an understanding of its angular structure will be essential for making use of existing predictions for the angular power spectrum of the EGRB\@.  

In this paper I calculate the angular power spectrum of gamma-ray emission from unresolved Galactic dark matter substructure under various assumptions about the properties of the subhalo population.  The terms `substructure' and `subhalos' are used here interchangeably to refer to dark matter clumps within a larger halo. 
I introduce the formulae for the gamma-ray emission from dark matter annihilation in \S\ref{sec:dmgamma}, and present the models used to describe the subhalo population in \S\ref{sec:dmmodel}.
The angular distribution of the emission from substructure relative to the smooth halo emission is discussed in \S\ref{sec:maps}.    In \S\ref{sec:powerspect} I present the angular power spectrum of gamma-ray emission from substructure and evaluate the prospects for detecting this signal with Fermi.  I discuss the implications of my results for indirect detection of Galactic dark matter in \S\ref{sec:discussion}, and summarize my conclusions in \S\ref{sec:summary}.

\section{Gamma-ray emission from dark matter annihilation}
\label{sec:dmgamma}

The intensity of gamma-ray emission from dark matter annihilation is given by
\begin{equation}
\label{eq:losintens}
I(\psi)=\frac{K}{4\pi}\int_{los}{\rm d}s\,\, \rho^{2}(s,\psi),
\end{equation}
where the variable $s$ denotes distance in the line of sight direction $\psi$, where $\psi$ is a set of orientation angles, and $\rho(s,\psi)$ is the density of dark matter.  All of the parameters which depend on the intrinsic properties of the assumed dark matter particle enter via the multiplicative factor $K$, defined by
\begin{equation}
\label{eq:kparticle}
K=\frac{N_{\gamma}(> \! E_{\rm th})}{2} \left(\frac{\langle\sigma v\rangle}{\mbox{\small cm}^{3}\, \mbox{\small s}^{-1}}\right) \left(\frac{\mbox{\small GeV}} {m_{\chi}}\right)^{\!\! 2},
\end{equation}
where $N_{\gamma} (> \!\! E_{\rm th})$ is 
the number of continuum photons produced per annihilation above the energy threshold $E_{\rm th}$, $\langle\sigma v\rangle$ is the thermally averaged annihilation cross-section times the relative particle velocity, and $m_{\chi}$ is the particle mass.  The factor of $1/2$ is appropriate for particles which are their own antiparticle, such as the neutralino and LKP\@. 
The analytic approximation to the continuum photon spectrum from neutralino annihilation from \citet{bergstrom_edsjo_ullio_01},
\begin{equation}
\label{eq:dnde}
\frac{{\rm d}N_{\gamma}}{{\rm d}x} = m_{\chi} \frac{{\rm d}N_{\gamma}}{{\rm d}E} = \frac{0.42 e^{-8x}}{x^{1.5}+0.00014},
\end{equation}
with $x \equiv E/m_{\chi}$, is adopted here to estimate $N_{\gamma} (> \!\! E_{\rm th})  \! = \! \int_{E_{\rm th}}^{m_{\chi}} {\rm d}E\, ({\rm d}N_{\gamma}/{\rm d}E)$.  When calculating the amplitude of the gamma-ray emission, I take the neutralino to be the dark matter particle, although my results can be easily extended to other WIMP candidates such as the LKP by substituting an appropriate photon spectrum. 
I set $\langle\sigma v\rangle  \! = \! 3 \! \times \! 10^{-26}$ cm$^{3}$ s$^{-1}$, the approximate value required for the assumed WIMP to account for the observed dark matter density \cite{jungman_kamionkowski_griest_96}, and optimistically choose $m_{\chi} \! = \! 85$ GeV, the value which maximizes $K$ for an energy threshold of $E_{\rm th} \! = \! 10$ GeV, appropriate for observations by Fermi.  The sensitivity of observations to other values of $m_{\chi}$ is discussed in \S\ref{sec:powerspect}.       

Clearly, the value of $K$ affects only the overall magnitude and not the spatial distribution of the gamma-ray emission.  As previously noted, the uncertainty in $K$ makes it difficult to constrain dark matter models using the amplitude of measured gamma-ray emission.  In contrast, the angular power spectrum from dark matter substructure is determined exclusively by the source distribution, so constraints from a measurement of the angular power spectrum are fairly insensitive to the uncertainties in the intrinsic properties of the dark matter particle.  For this reason, intensity is quoted here in units of $10^{30}$ $K^{-1}$ cm$^{-2}$ s$^{-1}$ sr$^{-1}$, which enables the results to be generalized to an arbitrary value of $K$\@.  For the 10 GeV energy threshold considered here, the most optimistic value of $K$ is $\sim10^{-30}$, which is the motivation for the prefactor $10^{30}$ in the chosen units.  The overall amplitude of the emission (determined by both the dark matter distribution via $\rho^{2}$ and also by the value of $K$) is, of course, a necessary consideration for determining the detectability of the signal, and is discussed in \S\ref{sec:powerspect}.

\section{The Galactic dark matter distribution}
\label{sec:dmmodel}

Analytic models \citep[e.g.,][]{sheth_tormen_99, seljak_00, cooray_sheth_02} can accurately predict the large scale clustering properties of dark matter, and are a natural choice for calculating the angular power spectrum of gamma-ray emission from large scale structure \citep[as in][]{ando_komatsu_06, cuoco_hannestad_haugbolle_etal_07}.
However, the applicability of these models on subgalactic scales is limited by non-linear effects, such as tidal stripping of the subhalos.  Moreover, for the case of emission from Galactic dark matter, our position in the halo must be taken into account since the source density varies significantly over the halo volume.  In light of these considerations, I use the results of numerical simulations to model the Galactic dark matter distribution.

\subsection{Halo properties and conventions}

The virial radius of a halo is defined as the radius within which the mean enclosed density is $\Delta \times \rho_{\rm cr}$, where $\rho_{\rm cr} \! = \! 3 H_{0}^{2}/8 \pi G$ is the critical density for closure at $z=0$. 
Several choices for $\Delta$ are used in the literature, resulting in slight variations in the definition of virial radius and, by extension, quantities defined in relation to the virial radius.  Following the convention used in \cite{gao_navarro_cole_etal_07}, I set $\Delta \! = \! 200$, which defines the virial mass $M_{200}$ and virial radius $r_{200}$ of a halo via $M_{200} \! = \! \frac{4}{3} \pi r_{200}^{3} \Delta \rho_{\rm cr}$. 
For the halo density profiles considered here, the concentration of a halo is defined by $c_{200} \equiv r_{200}/r_{s}$, where $r_{s}$ is the scale radius of the profile.

\subsection{The subhalo radial distribution}

Numerical simulations agree that the distribution of subhalos is \emph{anti-biased} relative to the smooth component of the host mass distribution,  and convergence studies have demonstrated that the consistent lack of subhalos in the central regions of the halo is not due to overmerging or numerical limitations, although the extent of the bias can be influenced by subhalo selection criteria  \cite{de-lucia_kauffmann_springel_etal_04, diemand_moore_stadel_04, gao_white_jenkins_etal_04, nagai_kravtsov_05, shaw_weller_ostriker_etal_07}.  Tidal mass loss has often been invoked to explain the reduction in subhalos near the center of the host halo, but the degree to which subhalos with masses much smaller than the resolution limit of the simulation would be disrupted is unknown.  

Many previous studies of the substructure annihilation flux made the more optimistic assumption that the subhalo distribution is \emph{unbiased} with respect to the smooth dark matter component (i.e., the subhalos trace the mass distribution of the host halo) \citep[e.g.,][]{bergstrom_edsjo_gondolo_etal_99, calcaneo-roldan_moore_00, berezinsky_dokuchaev_eroshenko_03}.  Two exceptions are recent work by \citet{pieri_bertone_branchini_08} who modified this assumption to account for tidal disruption near the center of the host by introducing a minimum radius dependent on the mass of the subhalo, and \citet{kuhlen_diemand_madau_08} who utilize an anti-biased subhalo distribution.  
An important consequence of an anti-biased subhalo radial distribution is that the typical distances of subhalos from our position are much larger than those of an unbiased distribution, and as such the typical subhalo fluxes are correspondingly smaller.  It is reasonable to expect that the subhalo radial distribution may thus affect the detectability of the Galactic dark matter signal, and it is shown here that an anti-biased distribution does indeed result in a considerable decrease in the total flux.

I examine both an unbiased and an anti-biased distribution, with the expectation that the true subhalo radial distribution lies somewhere in between. 
The unbiased distribution, while probably unrealistic for subhalos of moderate masses, may more accurately describe the distribution of the far more numerous low-mass subhalos; using this distribution also enables comparison with previous work.  The anti-biased distribution represents a more pessimistic scenario for detecting the annihilation flux, but is well-motivated by numerical work and dynamical arguments.  

I first consider the scenario in which the subhalo radial distribution is unbiased with respect to the smooth component, and model the smooth component using the density profile proposed by \citet*[hereafter NFW]{navarro_frenk_white_95}.  The NFW profile is given by
\begin{equation}
\label{eq:nfwden}
\rho_{\rm \scriptscriptstyle NFW}(r)=\frac{\rho_{s,\rm \scriptscriptstyle NFW}}{x(1+x)^{2}}
\end{equation}
with $x \equiv r/r_{s}$, where $r_{s}$ is a scale radius and $\rho_{s,\rm \scriptscriptstyle NFW}$ is a characteristic density.
For a host halo described by a NFW density profile, the cumulative fraction of subhalos within $x$ is
\begin{equation}
\frac{N_{\rm \scriptscriptstyle NFW}(< \! x)}{N_{\rm tot}}=\frac{f(x)}{f(c_{200})},
\end{equation}
where $f(x) \! = \! {\rm ln}(1+x)-(x/(1+x))$, $c_{200}$ is the host halo concentration, and $N_{\rm tot}$ is the total number of subhalos within $r_{200}$.  I subsequently refer to this as the unbiased radial distribution.  
For the case in which the subhalo distribution is anti-biased relative to the smooth component, I use the fitting formula from \citet{gao_white_jenkins_etal_04} to describe the radial distribution of the subhalos.  The number of subhalos within $z$, $N_{\rm anti}(< \! z)$, is 
\begin{equation}
\label{eq:gao_rad_profile}
\frac{N_{\rm anti}(< \! z)}{N_{\rm tot}} = \frac{(1+ac_{200})z^{\beta}}{(1+ac_{200}z^{\gamma})},
\end{equation}
with $z \equiv r/r_{200}  \! = \!  x/c_{200}$, $a \! = \! 0.244$, $\beta  \! = \!  2.75$, $\gamma  \! = \!  2$, and the parameters $x$, $c_{200}$, and $N_{\rm tot}$ defined as in the NFW radial distribution.  I refer to this as the anti-biased radial distribution.  These two radial distributions are implemented using the structural parameters for the Milky Way halo from \cite{klypin_zhao_somerville_02}: $r_{s} \! = \! 21.5$ kpc and $c_{\rm vir} \! = \! 12$ for a NFW density profile.\footnote{The definition of $r_{\rm vir}$ in \cite{klypin_zhao_somerville_02} differs from that used here, but the difference between the Milky Way halo $c_{\rm vir}$ and the corresponding $c_{200}$ value is negligible and has no significant effect on the subhalo radial distribution, so here it is assumed that $c_{200} \! = \! c_{\rm vir}$.}  Spherical symmetry is assumed for both radial distributions.

\subsection{The subhalo mass function}

Simulations generally find that the cumulative mass function of subhalos (number of subhalos with mass greater than $M_{\rm sub}$) follows a power law, 
$N(> \! M_{\rm sub}) \propto M_{\rm  sub}^{-\alpha_{\rm m}}$,
with $\alpha_{\rm m}$ ranging from approximately 0.8 to 1.0 \cite{moore_ghigna_governato_etal_99, ghigna_moore_governato_etal_00, de-lucia_kauffmann_springel_etal_04, gao_white_jenkins_etal_04, diemand_kuhlen_madau_07a, diemand_kuhlen_madau_07b, shaw_weller_ostriker_etal_07}.  
The highest resolution simulations to date resolve subhalos with $M_{\rm sub} \gtrsim 10^{6}$ M$_{\odot}$ at $z \! = \! 0$, although the mass function is generally assumed to hold for subhalo masses far below this limit.  The uncertainty in $\alpha_{\rm m}$ translates to substantial differences in the total number of subhalos when the mass function is extrapolated several orders of magnitude below the reach of simulations, so the three cases $\alpha_{\rm m}$ = 0.8, 0.9, and 1.0 are considered here to illustrate the sensitivity of the angular power spectrum to the slope of the subhalo mass function.  For each choice of $\alpha_{\rm m}$, the cumulative mass function is normalized at $M_{\rm sub} \! = \! 10^{8}$ M$_{\odot}$ to match the value obtained from equation (14) in \citep{diemand_kuhlen_madau_07b}:
\begin{equation}
\label{eq:vlmassfunc}
N(> \! M_{\rm sub}) = 0.0064 \left(M_{\rm  sub}/M_{\rm 200}\right)^{-\alpha_{\rm m}}
\end{equation}
with $\alpha_{\rm m} \! = \! 1$, setting the Milky Way halo mass $M_{\rm 200} \! = \!  10^{12}$ M$_{\odot}$\footnote{The mass of the halo in the Via Lactea simulation from which equation~(\ref{eq:vlmassfunc}) was determined is almost a factor of 2 greater than the Milky Way halo mass assumed here, and so for a host halo of that mass, equation~(\ref{eq:vlmassfunc}) predicts roughly twice as many subhalos above a given subhalo mass.}.
The value $\alpha_{\rm m} \! = \! 0.9$ is chosen as the fiducial case and, unless otherwise stated, all figures refer to realizations generated with this value.  Differences in the results for the other two choices of $\alpha_{\rm m}$ are discussed in the text.  The model adopted here does not include a treatment of sub-subhalos \citep{kuhlen_diemand_madau_08}, which would increase the total annihilation flux.

In this work, I focus on two possibilities for the minimum subhalo mass: $M_{\rm min}  \! = \!  10^{7}$ M$_{\odot}$ and $10$ M$_{\odot}$.  The $M_{\rm min}  \! = \!  10^{7}$ M$_{\odot}$ case is chosen to represent an upper limit on $M_{\rm min}$, and is motivated by the masses of the known Milky Way satellites \cite{strigari_bullock_kaplinghat_etal_07}.  The $M_{\rm min}  \! = \!  10$ M$_{\odot}$ case, on the other hand, is not intended as a lower limit on $M_{\rm min}$.  Instead, this value is chosen to be small enough to clearly demonstrate the impact of a large number of low-mass subhalos on the power spectrum, while still remaining computationally tractable for the approach used in this work.  Empirically, this value is also roughly the limit at which the typical angular separation of the subhalos falls below the angular resolution of current and upcoming experiments, and consequently the contribution from substructure below this mass limit appears as isotropic noise in the power spectrum.

\subsection{The subhalo density profile}

The NFW density profile has been widely used to model the dark matter distribution.  However, computational advances in recent years have enabled simulations to more accurately resolve the central structure of dark matter halos, and as a result, fits to a NFW density profile were shown to systematically deviate in the innermost regions \cite{navarro_hayashi_power_etal_04, merritt_graham_moore_etal_06, gao_navarro_cole_etal_07}.
A density profile with the logarithmic slope a power law of radius, such as that introduced by \citet{einasto_65} to fit stellar density profiles, was shown to better fit simulation data over a variety of mass scales.
The Einasto profile is given by
\begin{equation}
\label{eq:einden}
\rho_{\rm \scriptscriptstyle Ein}(r) = \rho_{s,\rm \scriptscriptstyle Ein} \, \exp\left(\frac{2}{\alpha}\right) \, \rm{exp}\!\left[-\frac{2}{\alpha}\left(\frac{r}{r_{s}}\right)^{\! \! \alpha}\right],
\end{equation}
where $r_{s}$ and $\rho_{s,\rm \scriptscriptstyle Ein}$ are a scale radius and characteristic density.  Note that the parameter $\alpha$ in this formula differs from the parameter $\alpha_{\rm m}$ defined earlier to describe the slope of the subhalo mass function.
In \citet{gao_navarro_cole_etal_07},  $\alpha \sim 0.16$ was found to give accurate fits to the density profiles of all but very massive halos, so I adopt this value for the subhalo population. 

Since the annihilation rate scales as the density squared, the majority of the flux originates from the central regions of the halo, and consequently both the radial profile and the amplitude of the emission are quite sensitive to the assumed inner density profile.  
Most previous work on indirect detection of dark matter has adopted the NFW profile or in some cases the steeper Moore et al.\ profile \cite{moore_governato_quinn_etal_98}.  For a given halo mass and concentration, the Einasto profile generally results in a factor of a few enhancement in the flux compared with the NFW profile, with most of the increase concentrated in the central regions.
For reference, a more detailed comparison is presented in the Appendix.

Several studies have found that the concentration and mass of a halo are correlated \cite{bullock_kolatt_sigad_etal_01, eke_navarro_steinmetz_01, maccio_dutton_van-den-bosch_etal_07, gao_navarro_cole_etal_07}, which allows the two parameter density profile adopted here to be entirely specified by the mass of the halo.  I use the relation given in \citet{gao_navarro_cole_etal_07} for halos fit with Einasto profiles, 
\begin{equation}
\label{eq:c200_m200_rel}
c_{200}(M_{200}) \propto M_{200}^{-0.138},
\end{equation}
the slope of which is consistent with that found by \citet{bullock_kolatt_sigad_etal_01} for isolated halos fit with NFW profiles.  Although \citeauthor{bullock_kolatt_sigad_etal_01}\ report a steeper dependence on mass for subhalos than for isolated halos, the work presented here requires extrapolation of the mass function to subhalo masses many orders of magnitude below those tested by simulations, so I \emph{conservatively} adopt the shallower slope of equation (\ref{eq:c200_m200_rel}) to determine the concentrations for subhalos.  This is a conservative choice because a steeper slope would imply much more concentrated halos at the smallest masses, and thus a larger annihilation signal.

I emphasize that the definition of subhalo mass is somewhat arbitrary, since tidal stripping removes the outer regions of the subhalo making the virial radius meaningless, and the background density of the host halo complicates the determination of the subhalo's radial extent.  For the purpose of determining the subhalo concentrations via equation (\ref{eq:c200_m200_rel}), I set the virial mass $M_{200} \! = \! M_{\rm sub}$.  
Subhalos surviving to the present day are likely to have suffered severe mass loss due to tidal forces, which preferentially strip the outer regions of the subhalo \cite{kazantzidis_mayer_mastropietro_etal_04}.  This process increases the concentration of stripped subhalos by truncating their radial extent.  However, the density profile in the innermost regions, from which the majority of the annihilation flux originates, is expected to remain intact, particularly in the case of subhalos at the low-mass end of the mass function which are more resilient due to their high concentrations.

\section{Gamma-rays from Galactic dark matter}
\label{sec:maps}

The Galactic dark matter signal includes emission from the smooth dark matter halo and from substructure.  The main focus of this work concerns small-scale anisotropies from Galactic substructure, and as such the smooth component is a foreground to the desired signal.  On the other hand, for a measurement of the large-scale dipole feature due to emission from the smooth halo, unresolved substructure may have an important effect on the properties of the predicted signal.  In this section I consider the spatial distribution of each component to assess the impact of the smooth component on the detectability of anisotropies from substructure, and comment in \S\ref{sec:discussion} on the implications of unresolved substructure for a measurement of the dipole.

\begin{figure}[ht]
	\centering
	\includegraphics[width=0.9\textwidth]{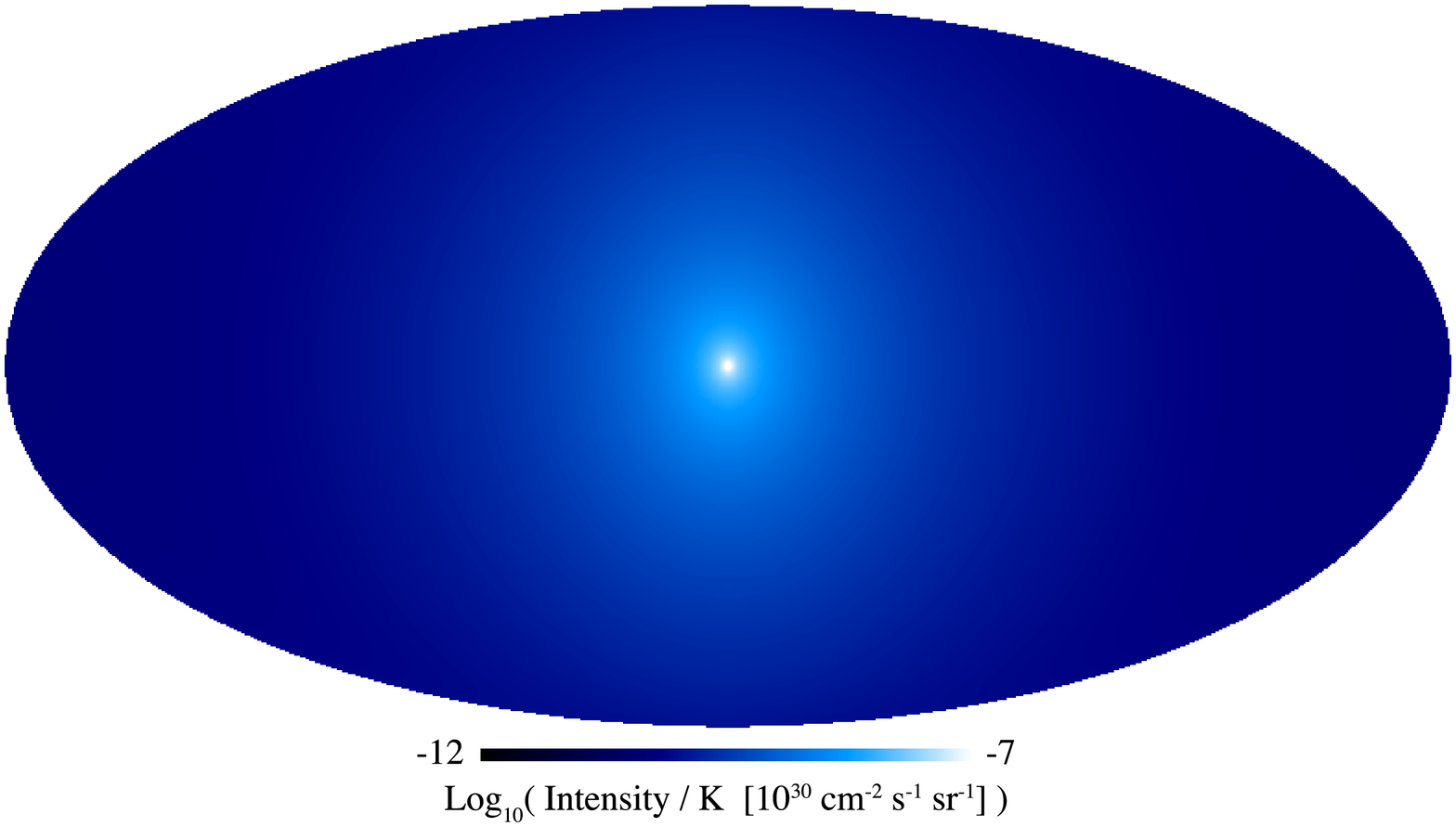}
	\includegraphics[width=0.9\textwidth]{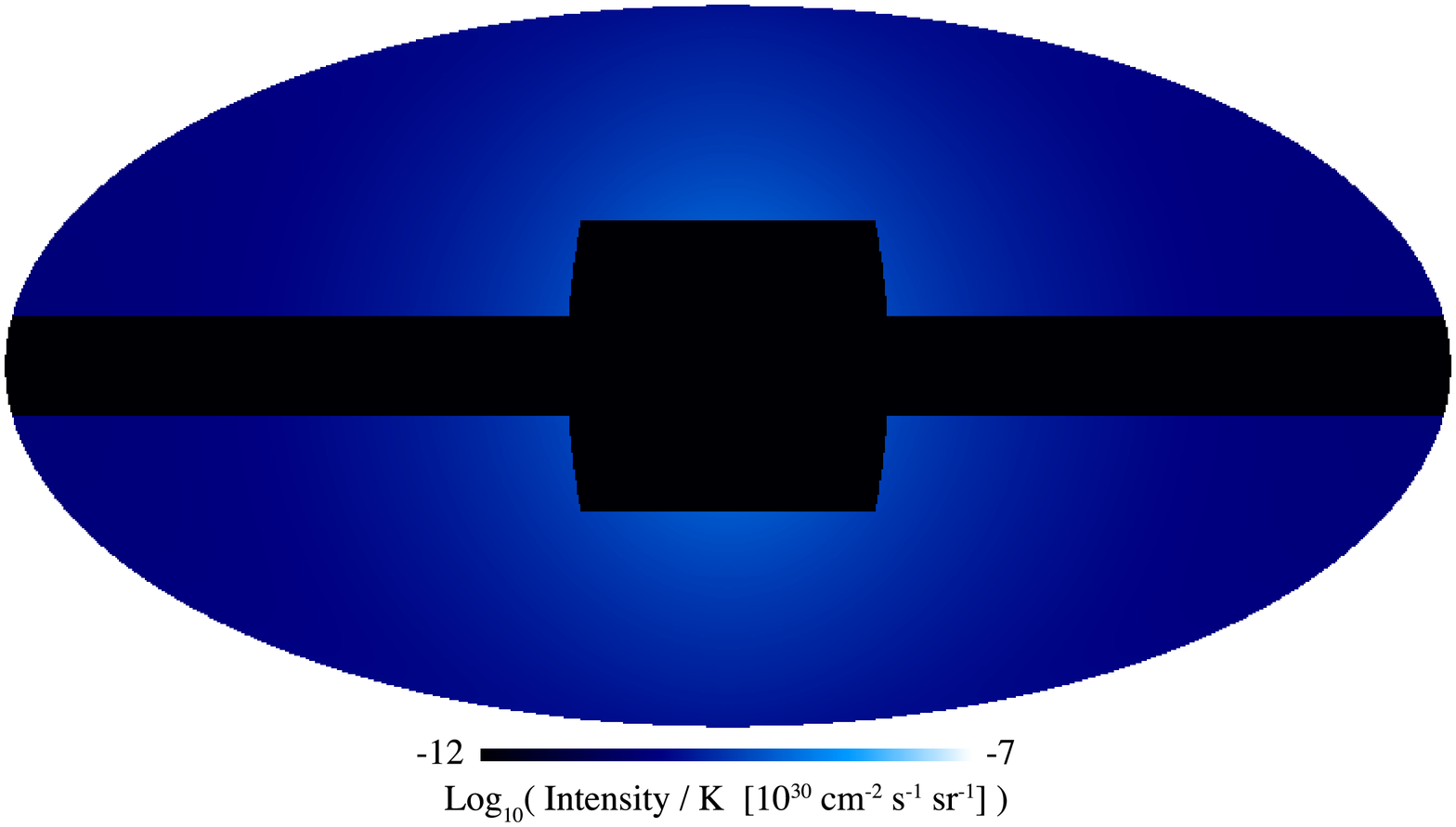}
	\caption{({\it Top panel:}) Gamma-ray intensity per $K$ from the dark matter halo without substructure (assuming the smooth component accounts for 100\% of the dark matter density) as observed from our position.  The map is centered on the Galactic Center.  ({\it Bottom panel:}) Same as top panel, with the Galactic emission mask of \cite{sreekumar_bertsch_dingus_etal_98} applied.
	\label{fig:smooth_halo}}
\end{figure}

I begin by calculating the gamma-ray emission from the smooth halo component.  The Milky Way halo is modeled with a NFW density profile described by the parameters given in \S\ref{sec:dmmodel}, and for this calculation all of the dark matter is assumed to be in the smooth component.  In the case that some fraction $f_{\rm sub}$ of Galactic dark matter is in the form of substructure, the amplitude and radial profile of the emission from the smooth component will be decreased according to $f_{\rm sub}$ and the subhalo radial distribution.  For the mass functions and minimum subhalo masses considered here, $f_{\rm sub}$ is at most~$\sim\! 15$\% and any change in the smooth component's emission due to some of the dark matter mass being in substructure is small in the cases of interest.  

The intensity per $K$ from the smooth halo as seen by an observer at 8.5 kpc from the Galactic Center is shown in the top panel of Figure~\ref{fig:smooth_halo} with a logarithmic color scale.  
The Galactic Center is a strong feature, several orders of magnitude brighter than most of the map.  However, astrophysical sources of gamma-rays near the Galactic Center are expected in most scenarios to dominate over the dark matter signal in that region; in fact, a recent analysis has shown that the signal-to-noise ratio for detecting the dark matter halo emission is optimized for an angular window about the Galactic Center of more than $10^{\circ}$ for the case of a NFW density profile \cite{serpico_zaharijas_08}.  
Data from regions of the sky expected to be highly contaminated, such as the Galactic Center, will not be useful for measuring the angular power spectrum of substructure emission due to low signal-to-noise.  As a rough indicator of the sky regions which are likely to be highly contaminated, the Galactic emission mask used in \cite{sreekumar_bertsch_dingus_etal_98} is shown overlaid on the halo emission map in the lower panel of Figure~\ref{fig:smooth_halo}.  This mask excludes the region around the Galactic Center ($|b| \! < \! 30^{\circ}$ for $| \ell | \! < \! 40^{\circ}$) and the Galactic plane ($|b| \! < \! 10^{\circ}$ for all $\ell$).  Here $\ell$ refers to Galactic longitude, but throughout the remainder of the paper $\ell$ is used in the standard notation as an index of the power spectrum coefficients.  This mask largely removes the dipole feature, and reduces the mean intensity of the halo emission by roughly a factor of 2.  

\begin{figure}[ht]
	\centering
	\includegraphics[width=0.9\textwidth]{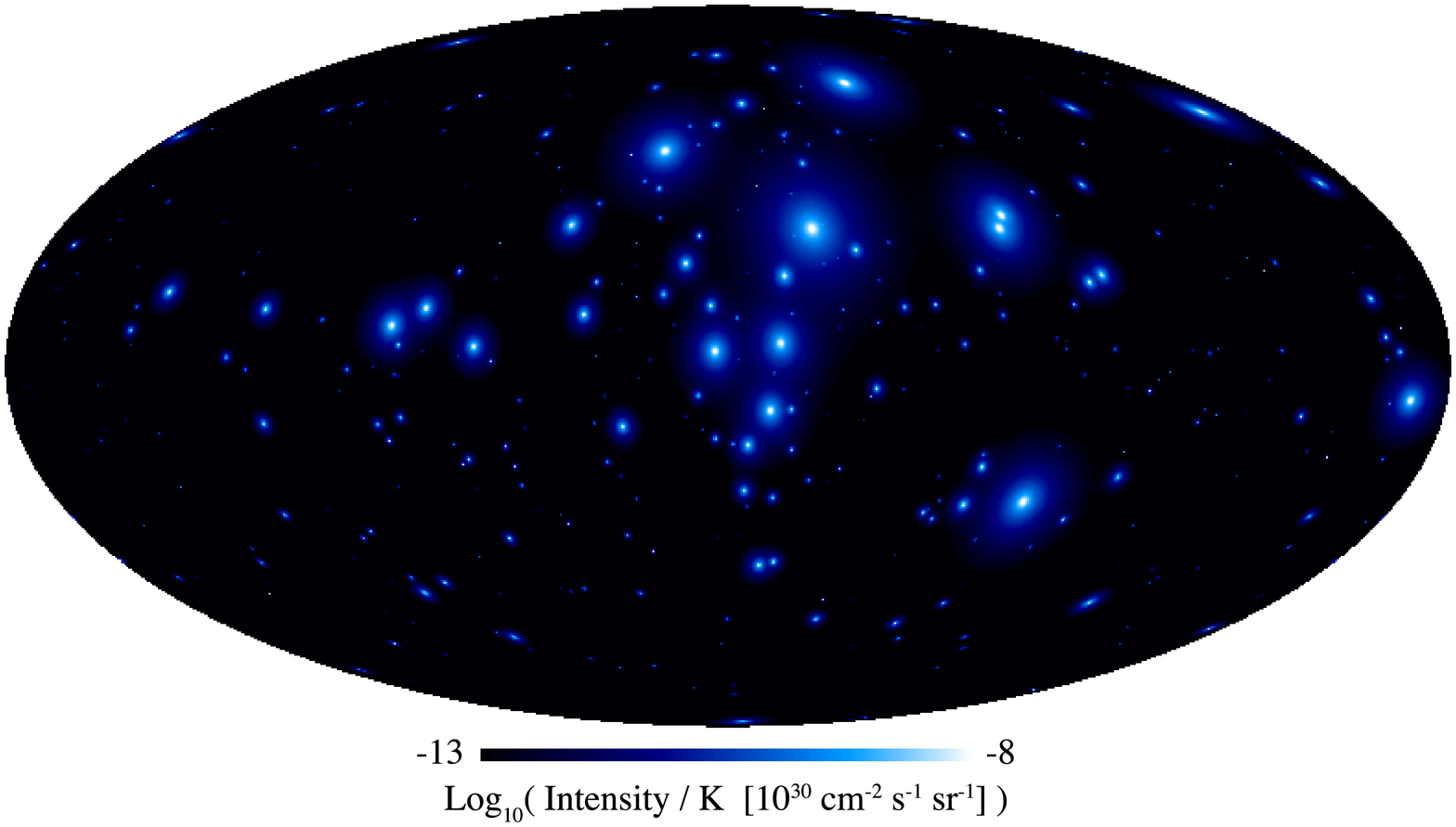}
	\includegraphics[width=0.9\textwidth]{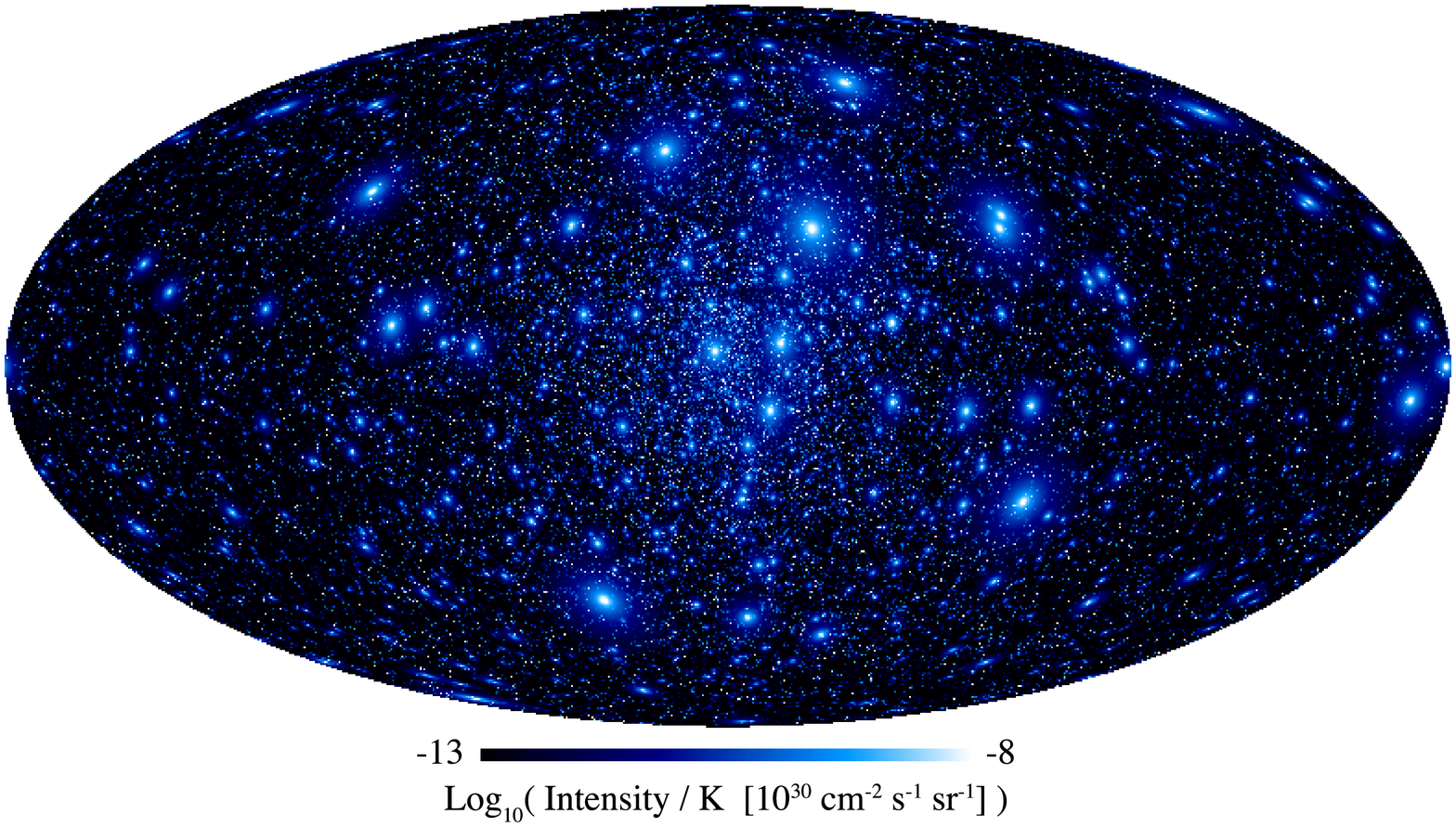}
	\caption{Gamma-ray intensity per $K$ from substructure for the unbiased radial distribution with $\alpha_{m} \! = \! 0.9$ and minimum subhalo masses of $M_{\rm min} \! = \! 10^{7}$ M$_{\odot}$ ({\it top}) and 10 M$_{\odot}$ ({\it bottom}).  Clustering of the subhalos in the direction of the Galactic Center (the center of the map) is quite pronounced, particularly in the $M_{\rm min} \! = \! 10$ M$_{\odot}$ case.
	\label{fig:map_nfw}}
\end{figure}

\begin{figure}[ht]
	\centering
	\includegraphics[width=0.9\textwidth]{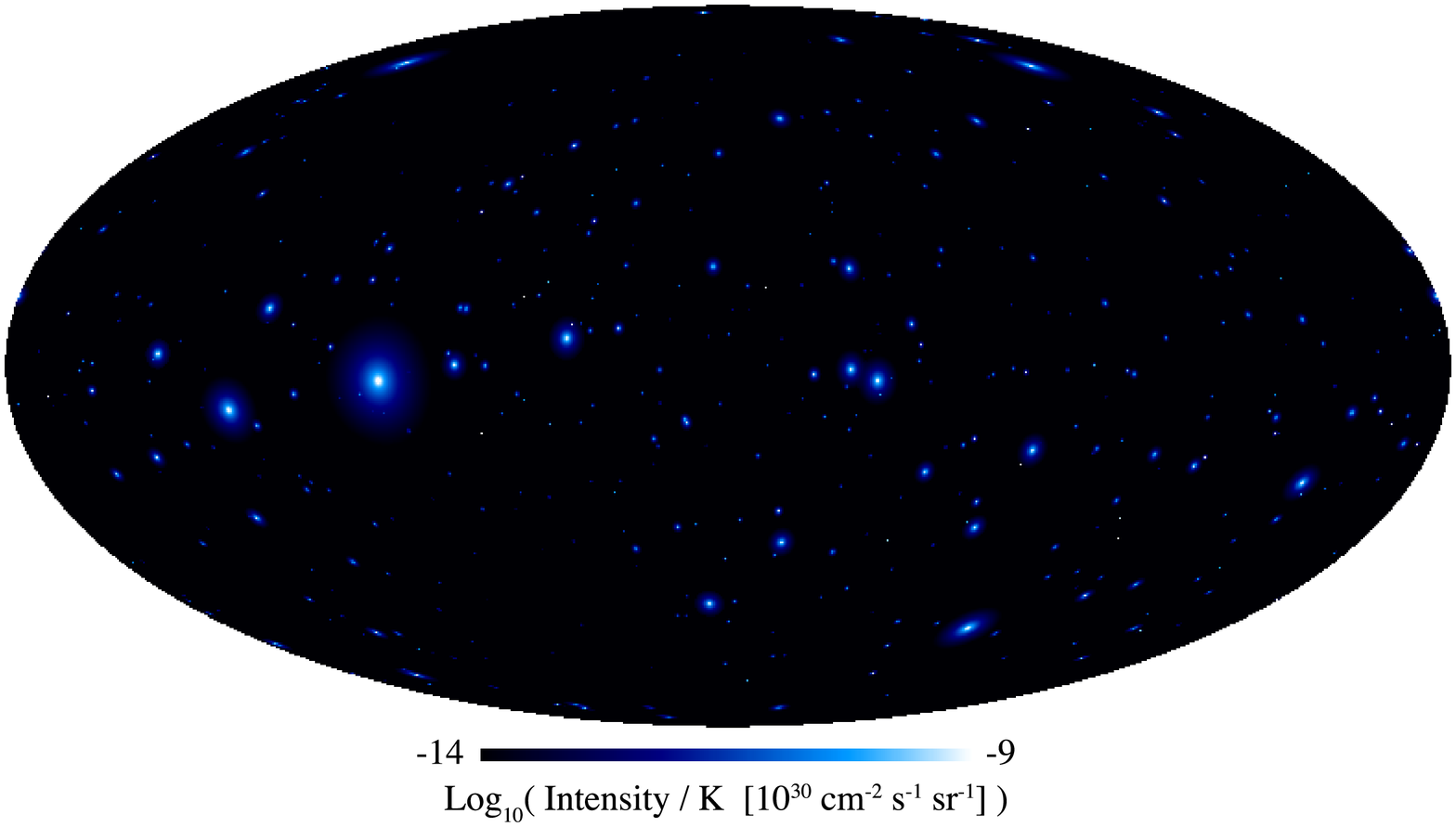}
	\includegraphics[width=0.9\textwidth]{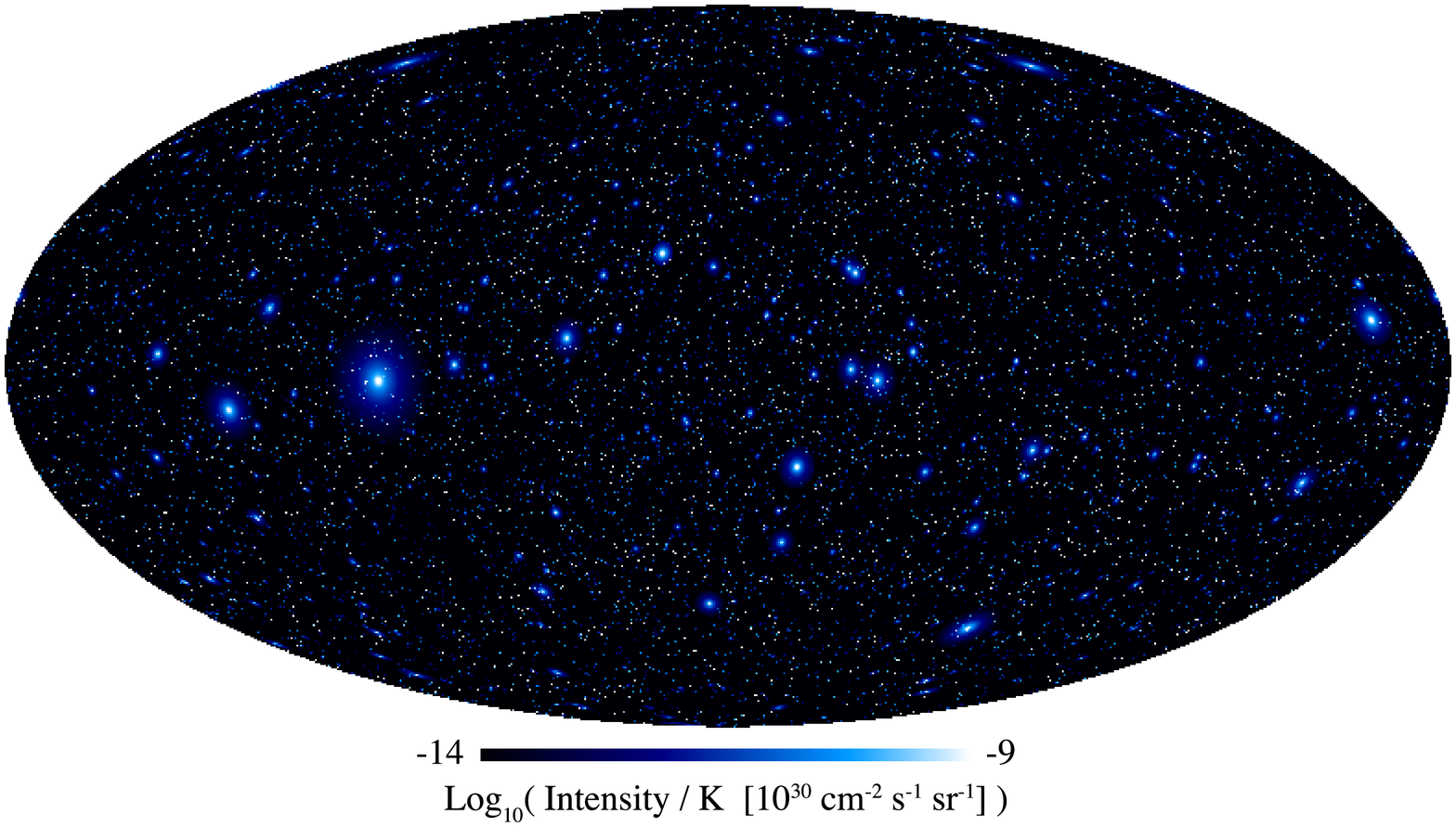}
	\caption{Gamma-ray intensity maps as in Figure~\ref{fig:map_nfw} for realizations of substructure in the anti-biased radial distribution.  The subhalos in these maps are typically much fainter than those in Figure~\ref{fig:map_nfw}: note that the color scale for these maps is shifted down one order of magnitude relative to the scale in Figure~\ref{fig:map_nfw}.  In contrast to the maps of the unbiased radial distribution, the subhalos appear roughly isotropic.  
		\label{fig:map_gao}}
\end{figure}

I now consider the emission from Galactic substructure.  Realizations of the substructure component are generated using the models described in \S\ref{sec:dmmodel} for different choices of the slope of the mass function $\alpha_{\rm m}$ and the radial distribution.  The gamma-ray emission from substructure as seen by an observer at 8.5 kpc from the Galactic Center is shown in Figure~\ref{fig:map_nfw} for the unbiased radial distribution and in Figure~\ref{fig:map_gao} for the anti-biased radial distribution.   The realizations shown were generated using the fiducial value $\alpha_{\rm m} \! = \! 0.9$ extrapolated to minimum subhalo masses of $10^{7}$ M$_{\odot}$ (upper panels) and 10 M$_{\odot}$ (lower panels).  For the unbiased radial distribution a concentration of subhalos in the direction of the Galactic Center is apparent for both minimum subhalo masses, while for the anti-biased case the subhalos appear isotropic.  The anti-biased radial distribution also results in roughly an order of magnitude less total flux for a given minimum subhalo mass since the typical distances of subhalos are much larger than in the unbiased distribution (note that the color scale differs between the maps for the unbiased and anti-biased distributions).  The maps shown here are for the $\alpha_{\rm m} \! = \! 0.9$ mass function, but similar trends in the subhalo angular distribution and total flux are observed in realizations using $\alpha_{\rm m} \! = \! 0.8$ and 1.

\begin{figure}[ht]
	\centering
	\includegraphics[width=0.9\textwidth]{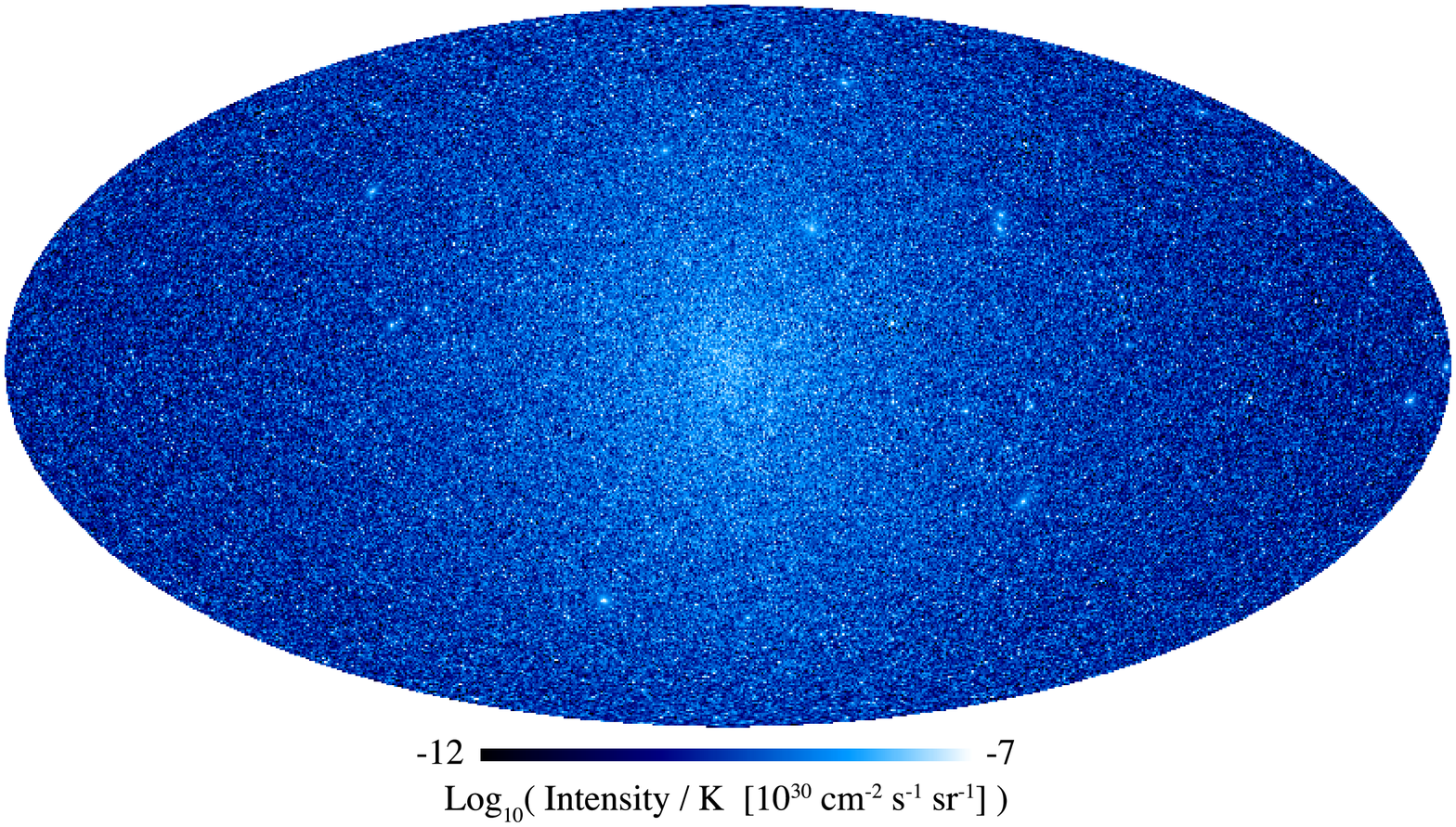}
	\includegraphics[width=0.9\textwidth]{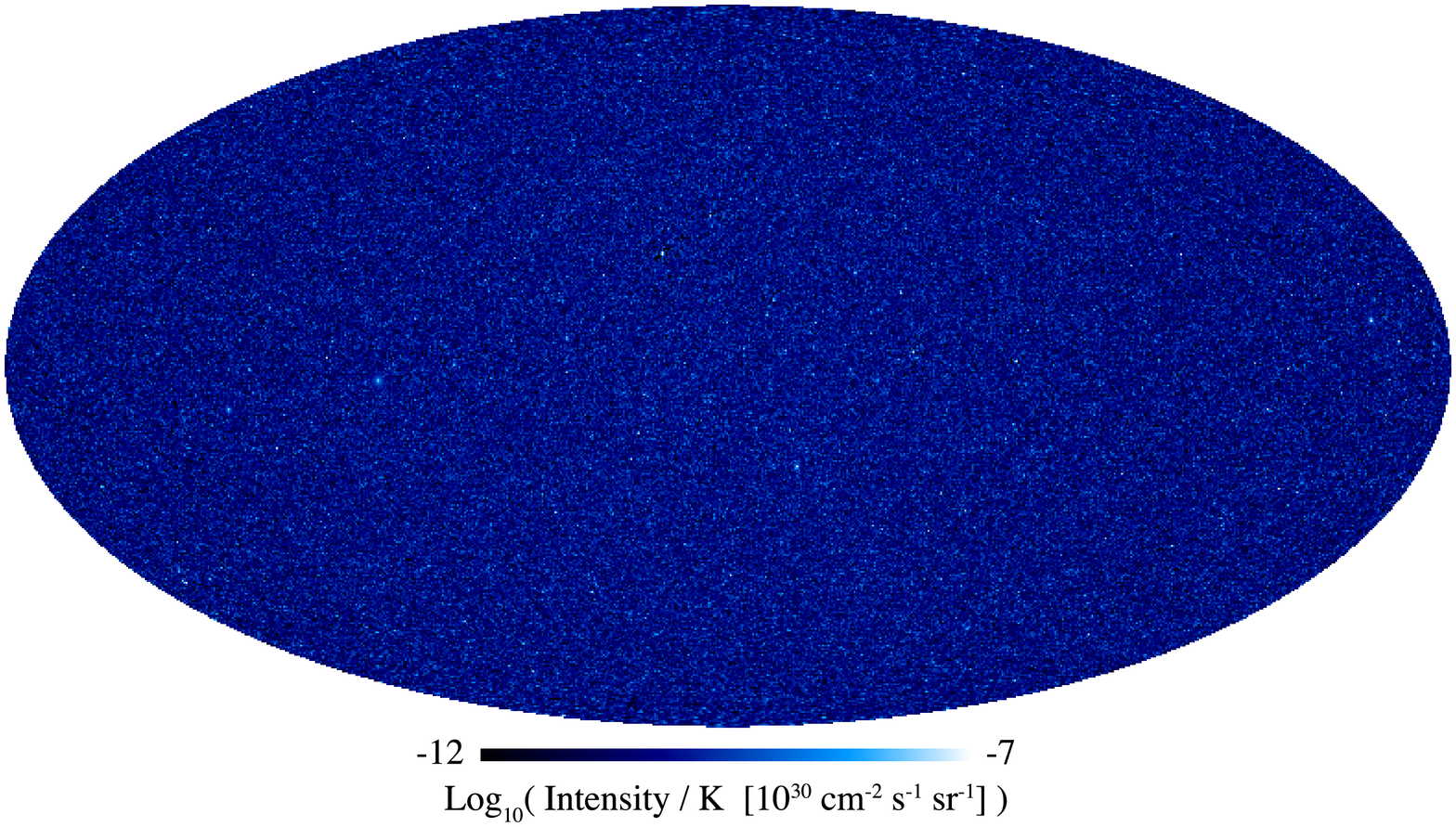}
	\caption{Maps of gamma-ray emission from substructure with $M_{\rm min} \! = \! 10$ M$_{\odot}$ smoothed with a Gaussian beam of width $\sigma_{b} \! = \! 0.1^{\circ}$. ({\it Top panel:}) Unbiased radial distribution (same realization as bottom panel of Figure~\ref{fig:map_nfw}).  ({\it Bottom panel:}) Anti-biased radial distribution (same realization as bottom panel of Figure~\ref{fig:map_gao}). 
	\label{fig:smooth_subs}}
\end{figure}

The vast majority of the subhalos in these realizations are not bright enough to be detected as point sources by Fermi for even the most optimistic particle physics scenarios \citep[see also][]{pieri_bertone_branchini_08, kuhlen_diemand_madau_08}, and instead contribute to the measured diffuse background.  The maps in Figure~\ref{fig:smooth_subs} simulate the effect of observing the subhalo emission from a $M_{\rm min} \! = \! 10$ M$_{\odot}$ scenario with an experiment having angular resolution comparable to that of Fermi.  The maps shown are those from the bottom panels of Figures~\ref{fig:map_nfw} and \ref{fig:map_gao} convolved with a Gaussian beam of width $\sigma_{\rm b} \! = \! 0.1^{\circ}$, Fermi's target angular resolution for $E \! > \! 10$ GeV\@.  The color scale is the same for both panels of this figure and matches that of the smooth halo emission map in Figure~\ref{fig:smooth_halo}, but is offset slightly from the scales used in the original maps in Figures \ref{fig:map_nfw} and \ref{fig:map_gao}.  For the unbiased radial distribution, a strong dipole feature due to the clustering of the subhalos near the Galactic Center is immediately apparent, mimicking that of the smooth halo emission, but for the anti-biased distribution the emission is remarkably isotropic.  Similar characteristics are observed for $M_{\rm min} \! = \! 10$ M$_{\odot}$ for all three choices of $\alpha_{\rm m}$.  
In contrast, convolving the maps for realizations with $M_{\rm min} \! = \! 10^{7}$ M$_{\odot}$ with the same Gaussian beam has no obvious effect on the emission maps (not shown), and the dark regions of the sky remain dark.

How does the smooth halo emission compare to the signal from substructure? 
Considering first the $\alpha_{\rm m} \! = \! 0.9$ case with $M_{\rm min} \! = \! 10$ M$_{\odot}$ shown here, for the unbiased radial distribution the intensity of the substructure near the Galactic Center is comparable to that of the smooth halo. 
In regions beyond a few tens of degrees from the Galactic Center, the mean map intensity for the unbiased subhalo distribution exceeds that of the smooth halo by more than an order of magnitude, and for the anti-biased distribution the flux is of roughly the same intensity as the mean halo emission in those regions.  If the subhalo mass function extends to masses several orders of magnitude lower than the 10 M$_{\odot}$ case shown here, as predicted for CDM, the diffuse emission for both radial distributions would be enhanced.  This possibility is discussed in \S\ref{sec:powerspect}.

This comparison with the smooth halo emission can be intuitively extended to other choices for the subhalo model.  Since for a given $M_{\rm min}$ a larger $\alpha_{\rm m}$ produces a realization with more subhalos, the mean map intensity increases or decreases according to the slope of the subhalo mass function.  Naturally, a larger $M_{\rm min}$ corresponds to fewer subhalos, and hence a smaller mean map intensity.  Of course, since these realizations are randomly generated from the models, statistical variation also affects the mean map intensities.  While generally small for realizations with $M_{\rm min} \! = \! 10$ M$_{\odot}$ due to the enormous number of subhalos present, this variation is predictably larger for the $M_{\rm min} \! = \! 10^{7}$ M$_{\odot}$ realizations.  (These trends in map intensity can also be seen in Figure~\ref{fig:cl_ratio_compare}, discussed in \S\ref{sec:powerspect}.)
Although the smooth halo emission would be in some cases a significant contaminant of the overall emission, it does not introduce small-scale fluctuations and thus primarily affects the measured power spectrum at the relevant angular scales by contributing noise.

\section{The angular power spectrum}
\label{sec:powerspect}

I calculate the angular power spectrum of gamma-ray emission using the HEALPix\footnote{http://healpix.jpl.nasa.gov} package \cite{gorski_hivon_banday_etal_05}. 
The emission maps are generated using the HEALPix resolution parameter $N_{\rm side}  \! = \!  4096$ which corresponds to a map with $N_{\rm pix}  \! = \!  12 N_{\rm side}^{2} \simeq 2 \! \times \! 10^{8}$ pixels, and angular resolution $\Theta_{\rm pix}  \! = \!  0.0143^{\circ}$, almost an order of magnitude smaller than Fermi's target angular resolution for $E \! > \! 10$ GeV.   These parameters are chosen to ensure that the window function of the map does not affect the predicted angular power spectrum in the multipole range accessible to the experiment.

The dimensionless quantity $\delta I (\psi) \equiv (I(\psi) - \langle I \rangle)/\langle I \rangle$ is defined as a function on the sphere describing the fluctuation in intensity $I$ in a direction $\psi$, normalized to the mean intensity of the map $\langle I \rangle$.
The angular power spectrum of $\delta I (\psi)$ is given by the coefficients $C_{\ell} \! = \! \langle\, | a_{\ell m} |^{2} \rangle$, with the $a_{\ell m}$ determined by expanding $\delta I (\psi)$ in spherical harmonics, $\delta I (\psi)  \! = \!  \sum_{\ell,m} a_{\ell m} Y_{\ell m}(\psi)$.

The measured power spectrum $C_{\ell}$ is the sum of the power spectra of the signal $C_{\ell}^{\rm s}$ from Galactic dark matter substructure and the background $C_{\ell}^{\rm b}$, weighted by their relative intensities squared:
\begin{equation}
\label{eq:clsum}
\langle I_{\rm tot} \rangle^{2} C_{\ell}=\langle I_{\rm s} \rangle^{2} C_{\ell}^{\rm s} + \langle I_{\rm b} \rangle^{2} C_{\ell}^{\rm b},
\end{equation}
where $\langle I_{\rm tot} \rangle  \! = \!  \langle I_{\rm s} \rangle \! + \! \langle I_{\rm b} \rangle$. This relation assumes the signal and background are uncorrelated (i.e., $\langle | a_{\ell m}^{\rm s} \! + \! a_{\ell m}^{\rm b} |^{2} \rangle  \! = \!   \langle | a_{\ell m}^{\rm s} |^{2} \rangle \! + \! \langle | a_{\ell m}^{\rm b} |^{2} \rangle$).  
In the case considered here, $C_{\ell}^{\rm b}$ is the power spectrum of the extragalactic gamma-ray background and any sources of Galactic gamma-ray emission other than dark matter substructure.

\begin{figure}
\centering
\includegraphics[width=0.9\textwidth]{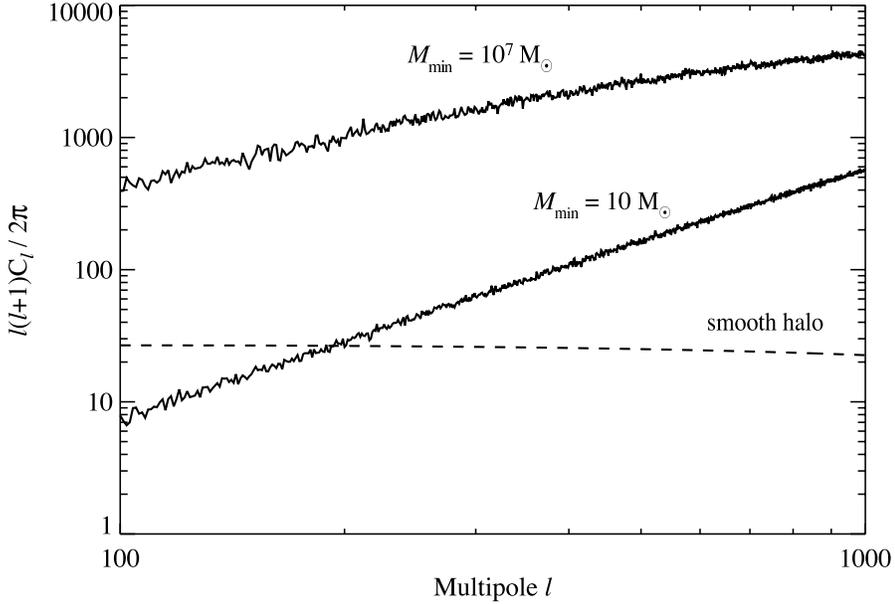}
\caption{Angular power spectrum of gamma-ray emission from dark matter substructure for minimum subhalo masses $M_{\rm min} \! = \! 10^{7}$ M$_{\odot}$ and 10 M$_{\odot}$ for the unbiased radial distribution and $\alpha_{m} \! = \! 0.9$.  The angular power spectrum of the smooth halo emission ({\it dashed line}) is shown for reference.  
\label{fig:powerspect}}
\end{figure}

The angular power spectrum of gamma-ray emission from Galactic substructure for a realization of the unbiased radial distribution with $\alpha_{\rm m} \! = \! 0.9$  is shown in Figure~\ref{fig:powerspect}.  Power spectra are plotted for $M_{\rm min} \! = \! 10^{7}$ M$_{\odot}$ and 10 M$_{\odot}$, and the power spectrum of the smooth halo emission is shown for comparison.  Each power spectrum is calculated from a map consisting exclusively of the emission from that source component.  The coefficients $C_{\ell}$ are dimensionless; each power spectrum should be multiplied by its $\langle I \rangle^{2}$ to compare the amplitudes in units of intensity squared.
For the $M_{\rm min} \! = \! 10^{7}$ M$_{\odot}$ case, the amplitude of the power spectrum is much greater than that for the $M_{\rm min} \! = \! 10$ M$_{\odot}$ case, and as expected, the smooth halo power spectrum has considerably less power at large multipoles than either of the substructure scenarios.  Since the Galactic Center region, where the smooth halo is brightest, will be highly contaminated, I note that the shape and amplitude of the \emph{measured} halo power spectrum may differ from that shown here, and the halo contribution to the measured power spectrum at these multipoles is likely to be smaller.  

The angular power spectra presented in this work are calculated from sky maps consisting entirely of `signal' emission, and thus represent the ideal (and unrealistic) case of perfect foreground cleaning.  Extracting the dark matter signal from a real data set will require careful treatment of Galactic foregrounds, the EGRB, and other contaminants \citep[see, e.g.,][]{de-boer_nordt_sander_etal_07}, and will introduce additional uncertainties in the measured power spectrum beyond the statistical errors considered below.  The error bars shown should therefore be regarded as the minimum expected uncertainty for a given scenario.

The $1$-$\sigma$ uncertainty in the measured power spectrum of the signal $C_{\ell}^{\rm s}$ is 
\begin{equation}
\delta C_{\ell}^{\rm s} = \sqrt{\frac{2}{(2\ell + 1)\,\Delta\ell\, f_{\rm sky}}} \left(C_{\ell}^{\rm s} + \frac{C_{\rm N}}{W_{\ell}^{2}}\right),
\end{equation}
for data in bins of width $\Delta\ell$, where $f_{\rm sky}$ is the fraction of the sky observed, $C_{\rm N}$ is the power spectrum of the noise, and $W_{\ell}  \! = \!  \exp(-\ell^{2}\sigma_{\rm b}^{2}/2)$ is the window function of a Gaussian beam of width $\sigma_{\rm b}$.
The noise power spectrum is the sum of the Poisson noise of the signal and the background, which takes the same value at all $\ell$:
\begin{equation}
C_{\rm N} = 4 \pi f_{\rm sky} \left(\frac{1}{N_{\rm s}} + \frac{(N_{\rm b}/N_{\rm s})^{2}}{N_{\rm b}}\right),
\end{equation}
where $N_{\rm s}$ and $N_{\rm b}$ are the number of signal and background photons, respectively.  
This simple estimate assumes a constant signal-to-noise for each pixel, and does not account for errors introduced by imperfect foreground cleaning or unequal weighting of pixels due to subtraction of spatially varying foregrounds or point sources.
I take $\Delta \ell \! = \! 100$, assume an energy threshold $E_{\rm th} \! = \! 10$ GeV, and use the following specifications for Fermi\footnote{http://fermi.gsfc.nasa.gov/science/}: effective area $A_{\rm eff} \! = \! 12000$ cm$^{2}$, field of view $\Omega_{\rm fov} \! = \! 2.4$ sr,  all-sky observing time $t_{\rm obs}^{\rm tot} \! = \! 5$ years, and $\sigma_{\rm b} \! = \! 0.1^{\circ}$.  This value of $\sigma_{\rm b}$ is appropriate for the chosen energy threshold of 10 GeV; at lower energies, Fermi has poorer angular resolution.  

Most observations with Fermi will be performed in sky scanning mode, resulting in fairly uniform exposure over the entire sky.  Since contamination of the dark matter signal in the Galactic plane is likely to be substantial, I approximate the effective amount of data available for this measurement by assuming that pixels falling within the Galactic emission mask of \cite{sreekumar_bertsch_dingus_etal_98} will have a prohibitively small signal-to-noise.  This leaves the usable fraction of the sky $f_{\rm sky} \! = \! 75$\%, so the effective observation time is $t_{\rm obs} \! = \! f_{\rm sky}t_{\rm obs}^{\rm tot}$ for a total all-sky observation time of $t_{\rm obs}^{\rm tot}$.  
Subtraction of bright point sources can also result in pixels with high noise levels which further reduce the usable $f_{\rm sky}$.  Predictions for the number of blazars Fermi will detect vary from $\sim3000$ to $\sim10000$ \cite{narumoto_totani_06}, but even in the latter case only an additional $\sim5$\% of the sky would be excluded, assuming each blazar contaminates pixels within an angular radius of $3 \sigma_{\rm b}$.

Fermi is expected to have excellent charged particle background rejection capabilities, reducing contamination of the high-latitude diffuse gamma-ray emission from all sources to less than 1\%.  The EGRB, extended Galactic diffuse emission, and the smooth dark matter halo, however, may contribute substantially to the measured emission.
   
 The EGRB is unfortunately extremely difficult to measure. Any experimental determination is heavily influenced by the adopted Galactic emission model, of which many of the input parameters are poorly constrained.  
The EGRET experiment measured diffuse emission from 30 MeV to a few tens of GeV, and produced an estimate of the amplitude and spectrum of the EGRB at those energies \cite{sreekumar_bertsch_dingus_etal_98}, but there is currently no consensus on the properties of the EGRB based on EGRET's measurement.  A subsequent analysis of the EGRET data by \citet{strong_moskalenko_reimer_04} suggested that the original determination overestimated the intensity of the EGRB is several energy bins, and \citet{stecker_hunter_kniffen_08} recently asserted that EGRET's measurement at energies above 1 GeV is unreliable due to a problem with the sensitivity calibration of the detector.  
Arguments in favor of a smaller EGRB than the EGRET estimate were also made by \citet{keshet_waxman_loeb_04}. 
Furthermore, with its improved sensitivity, Fermi is likely to resolve a large number of sources which for EGRET had contributed to the diffuse emission, but estimates for the reduction in the background vary \cite{narumoto_totani_06}.
For similar reasons, the intensity of the high-latitude Galactic diffuse emission above $\sim 10$ GeV is not well-known.  

Since the absolute intensity of the dark matter signal depends on the properties of the assumed dark matter particle, its amplitude relative to other sources of Galactic diffuse emission and the EGRB depend on the assumed particle properties.  The intensity of the substructure emission for a given substructure model relative to the smooth halo emission, on the other hand, is a more robust quantity. 

Although the amplitudes of the emission from dark matter substructure and likely backgrounds and foregrounds are uncertain, it is sensible to consider the possibility of moderate contamination of the signal from substructure.  To assess the impact of emission from sources other than substructure on the detectability of the signal, I calculate the measurement uncertainties for three reference cases:  $N_{\rm b}/N_{\rm s} \! = \!$ 0.01, 3, and 10.  The $N_{\rm b}/N_{\rm s} \! = \!$ 0.01 case represents the ideal scenario in which the dark matter signal is virtually free of contaminants, while the other two cases consider somewhat more realistic possibilities.  

\begin{figure}
\centering
\includegraphics[width=0.9\textwidth]{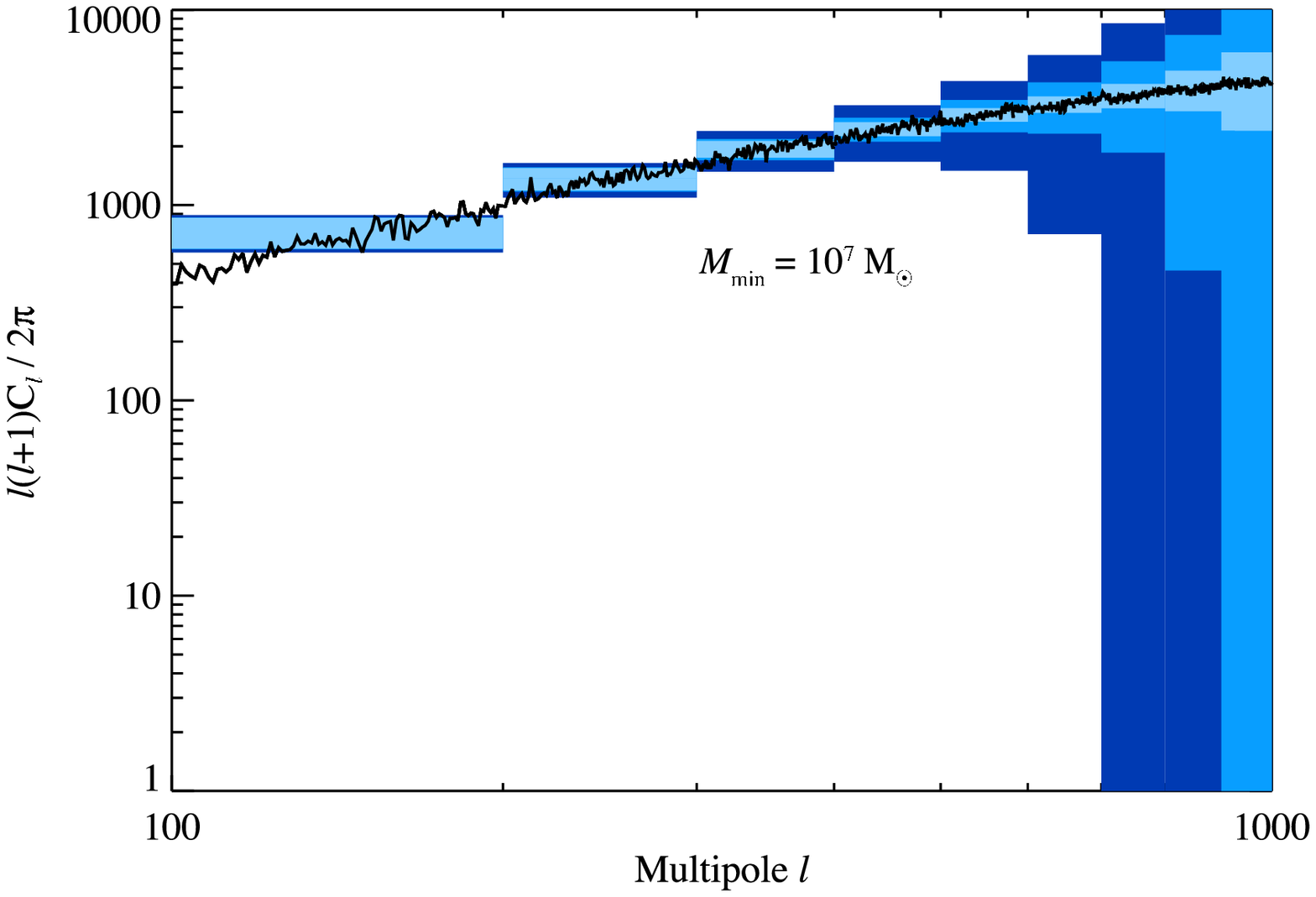}
\includegraphics[width=0.9\textwidth]{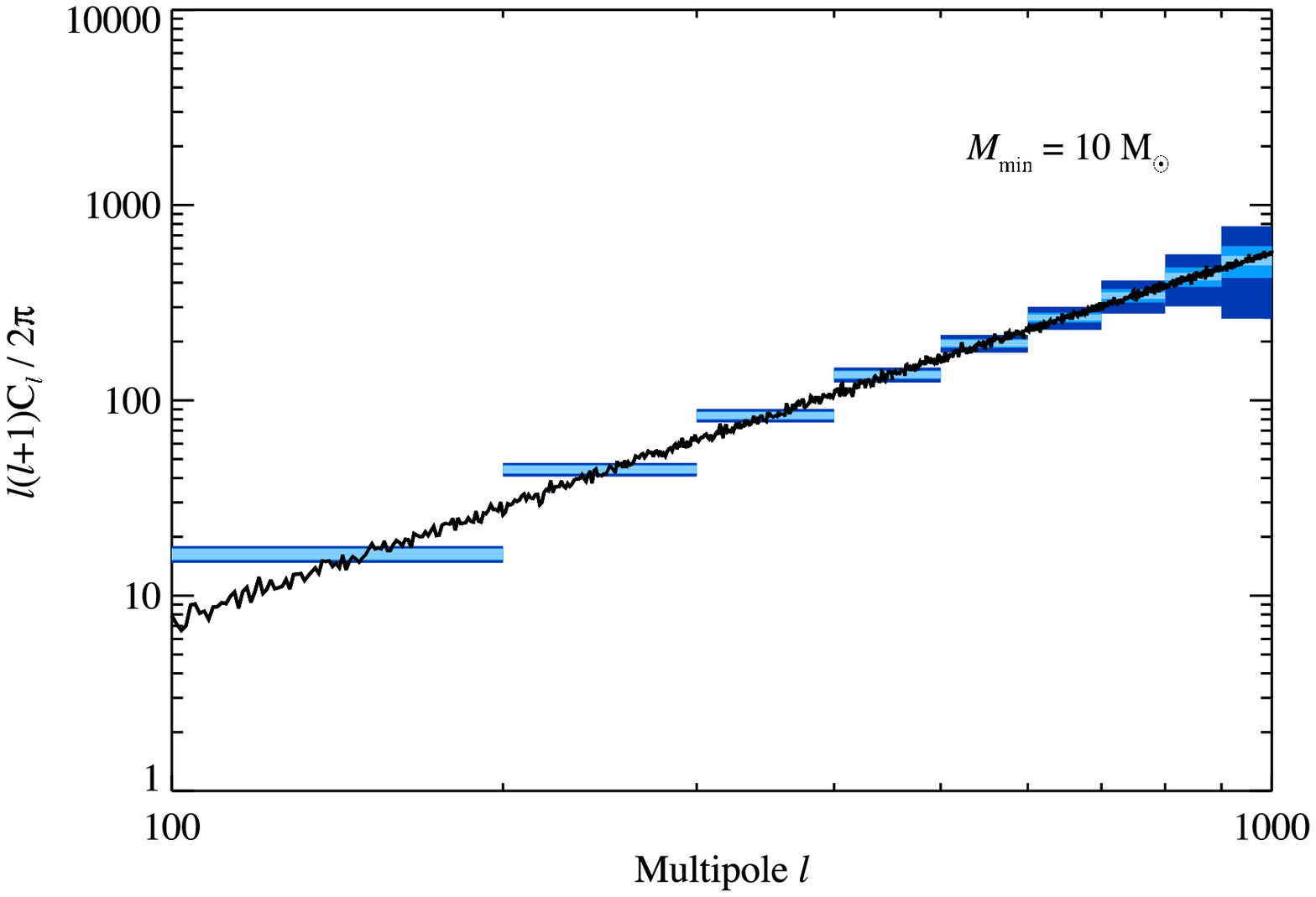}
\caption{Angular power spectra of gamma-ray emission from dark matter substructure as in Figure~\ref{fig:powerspect} for minimum subhalo masses $M_{\rm min} \! = \! 10^{7}$ M$_{\odot}$ ({\it top panel}) and 10 M$_{\odot}$ ({\it bottom panel}).  Error bars indicate the 1-$\sigma$ uncertainty in the measured power spectrum assuming 5 years of all-sky observation by Fermi, $f_{\rm sky} \! = \! 75$\%, and $N_{\rm b}/N_{\rm s} \! = \!$ 0.01 ({\it light blue}), 3 ({\it medium blue}), or 10 ({\it dark blue}).
\label{fig:powerspect_weighted}}
\end{figure}

The angular power spectrum of gamma-ray emission from Galactic substructure for 
the same realizations as in Figure~\ref{fig:powerspect} is shown in Figure~\ref{fig:powerspect_weighted} with expected measurement uncertainties for Fermi.  The $1$-$\sigma$ error bars are calculated for the three cases $N_{\rm b}/N_{\rm s} \! = \!$ 0.01, 3, and 10.  For the $M_{\rm min} \! = \! 10^{7}$ M$_{\odot}$ case, the error bars are larger than for the $M_{\rm min} \! = \! 10$ M$_{\odot}$ case due to the smaller number of signal photons.  Note that these error bars assume a fixed $N_{\rm b}/N_{\rm s}$, not a fixed $N_{\rm b}$, and $N_{\rm s}$ varies between realizations.
The amplitude of the predicted angular power spectrum of the EGRB due to dark matter annihilation $\ell(\ell+1)C_{\ell}/2\pi$ is less than $\sim 0.1$ out to $\ell \! = \! 1000$ \cite{ando_komatsu_06}, so the signal from Galactic substructure will dominate the power spectrum in this multipole range as long as the intensity of the EGRB does not exceed the signal intensity by more than $\sim 100$ for the $M_{\rm min} \! = \! 10^{7}$ M$_{\odot}$ scenario, or more than $\sim 10$ for the $M_{\rm min} \! = \! 10$ M$_{\odot}$ scenario.  However, contamination of the signal at the $N_{\rm b}/N_{\rm s} \sim 100$ level will clearly increase the error bars substantially, severely limiting the detectability of the signal.

\begin{figure}
\centering
\includegraphics[width=0.9\textwidth]{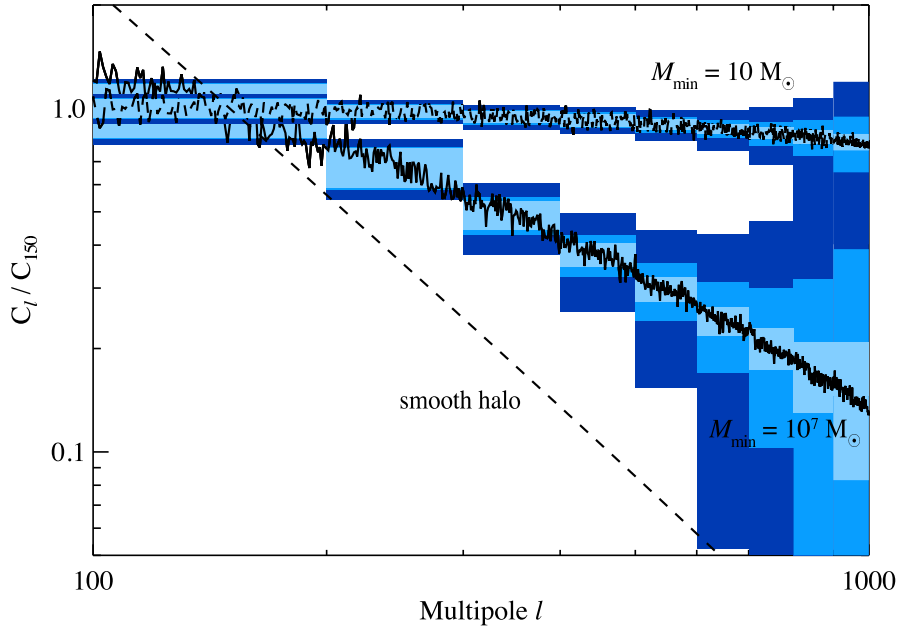}
\caption{Angular power spectra of gamma-ray emission from dark matter substructure as in Figure~\ref{fig:powerspect_weighted}.  To facilitate comparison, power spectra are normalized to $C_{150}$ and plotted without $\ell(\ell+1)/2\pi$ scaling.  The error bars overlap in the two bins with $\ell > 800$ for the $N_{\rm b}/N_{\rm s} \! = \! 10$ case ({\it dark blue}).  The smooth halo power spectrum ({\it dashed line}) is also shown. 
\label{fig:powerspect_raw}}
\end{figure}

Figure~\ref{fig:powerspect_raw} shows the power spectra and error bars for the same substructure scenarios as in Figure~\ref{fig:powerspect_weighted} but without scaling the coefficients $C_{\ell}$ by $\ell(\ell+1)/2\pi$; in the convention used in this figure, a noise power spectrum appears as a horizontal line.  The power spectra are also normalized to the value of each at $C_{150}$ to facilitate comparison of the spectral shapes. 
The power spectrum for $M_{\rm min} \! = \! 10^{7}$ M$_{\odot}$ falls off quickly with $\ell$, but for $M_{\rm min} \! = \! 10$ M$_{\odot}$ is almost flat (as a noise spectrum would be).  This characteristic difference in the shape of the power spectrum is present for the other cases of $\alpha_{\rm m}$, but with the spectrum falling off somewhat more steeply in the $\alpha_{\rm m} \! = \! 0.8$,  $M_{\rm min} \! = \! 10$ M$_{\odot}$ case than for other $\alpha_{\rm m}$ at this $M_{\rm min}$ due to the smaller total number of subhalos for this mass function slope.  The power spectra for the two minimum subhalo masses considered can be distinguished by Fermi over a large multipole range even with considerable contamination of the signal ($N_{\rm b}/N_{\rm s} \! = \! 10$).

For $100 \lesssim \ell \lesssim 700$, the noise spectrum $C_{\rm N}$ is the largest contributor to the error.  At smaller $\ell$ the uncertainties are dominated by cosmic variance, and at larger $\ell$ measurements are limited by the angular resolution of the experiment.  For fixed values of $\langle \sigma v \rangle$ and $E_{\rm th}$ and a specified dark matter distribution, the intensity of the gamma-ray emission (and hence $N_{\rm s}$) is determined by the dark matter particle mass $m_{\chi}$ via $K$\@.   Although the error bars shown in Figures~\ref{fig:powerspect_weighted} and \ref{fig:powerspect_raw} are for the optimistic case of $m_{\chi} \! = \! 85$ GeV, for a heavier particle with $m_{\chi} \! = \! 300$ GeV the number of signal photons is only a factor of~$\sim\! 2$ smaller, and the increase in the size of the error bars is just a factor of approximately $1.5$ to 2. 
The subhalo radial distribution, mass function, and minimum subhalo mass have a much larger influence on the measurement uncertainties due to the considerable spread in mean map intensities, which span several orders of magnitude for the models considered here and produce a similar spread in the size of the error bars.  

\begin{figure}
	\centering
	\includegraphics[width=0.9\textwidth]{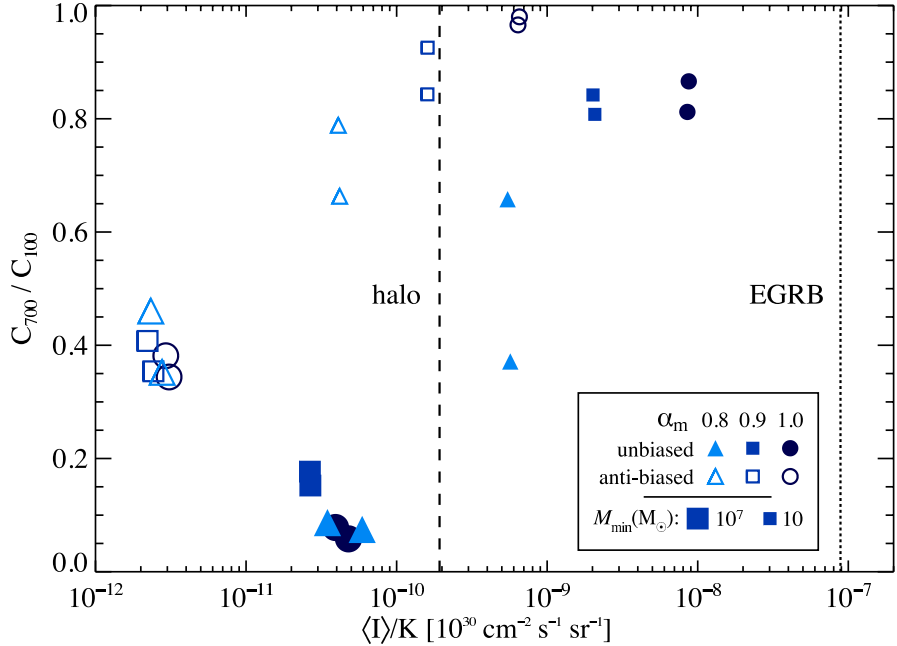}
	\caption{The ratio $C_{700}/C_{100}$ as a function of mean map intensity.  In this figure $C_{700}$ and $C_{100}$ are the coefficients $C_{\ell}$ averaged over the range $(\ell - 50)$ to $(\ell + 50)$.  Realizations generated using the unbiased (anti-biased) subhalo radial distributions are marked with filled (open) plot symbols.  The large (small) plot symbols denote realizations with $M_{\rm min} \! = \! 10^{7}$ M$_{\odot}$ (10 M$_{\odot}$).  Triangles ({\it light blue}), squares ({\it medium blue}), and circles ({\it dark blue}) correspond to $\alpha_{m}  \! = \! 0.8$, 0.9, and 1.0, respectively.  The mean intensity of the smooth halo ({\it dashed line}) and the EGRB intensity ({\it dotted line}) above 10 GeV from \cite{strong_moskalenko_reimer_04} for the adopted dark matter particle parameters are shown for reference.  
	\label{fig:cl_ratio_compare}}
\end{figure}

Figure~\ref{fig:cl_ratio_compare} illustrates the dependence of the mean map intensity, as well as the general shape of the angular power spectrum, on the assumed subhalo model.  In this figure the ratio of the amplitude of the angular power spectrum at $C_{700}$ and $C_{100}$ is used as a rough indicator of the shape of the power spectrum for the purpose of comparing the results of different subhalo models.  This ratio is not intended as a discriminator since a great deal of information contained in the power spectrum is lost by distilling the data to a single quantity.
Realizations generated with all three choices of $\alpha_{\rm m}$ for both radial distributions and both minimum subhalo masses are included in this figure.  
The mean intensity of the smooth halo (assuming that 100\% of the dark matter is in the smooth component),  $\langle I \rangle \! \sim \! 2 \! \times \!10^{-10}$ in units of $10^{30}$ $K^{-1}$ cm$^{-2}$ s$^{-1}$ sr$^{-1}$, is marked for reference.  This value is the mean of the all-sky emission, including the Galactic Center region.  As noted previously, the mean outside the Galactic emission mask of \cite{sreekumar_bertsch_dingus_etal_98} is roughly a factor of 2 smaller, and is a better estimate of the level of contamination of usable data.  For comparison, the EGRB intensity above 10 GeV from \cite{strong_moskalenko_reimer_04} is also marked, assuming the value of $K$ given by the dark matter particle parameters adopted.  Note that the intensity of the EGRB in the units shown is therefore model-dependent.

Trends in mean map intensity and power spectrum shape are evident from the clustering of the data points by shape (indicating mass function), size (indicating minimum subhalo mass), and whether open or filled (indicating radial distribution).   Statistical variations in the map intensity and power spectrum are also evident, although the number of realizations shown here is not sufficient to characterize the expected variation for different subhalo models.

This figure shows that the map intensity systematically increases with $\alpha_{\rm m}$ for each radial distribution for the 10 M$_{\odot}$ minimum mass, and that the unbiased radial distribution consistently produces a much larger mean map intensity than the anti-biased distribution for the same mass function and minimum subhalo mass.
The tendency for the power spectrum to decline at large multipoles in the $M_{\rm min} \! = \! 10^{7}$ M$_{\odot}$ scenarios can be seen by the low to moderate values of $C_{700}/C_{100}$ for these realizations.  In contrast, most of the realizations with the smaller minimum mass have ratios approaching unity.  A notable exception is one of the $M_{\rm min} \! = \! 10$ M$_{\odot}$ realizations for the unbiased radial distribution with $\alpha_{\rm m} \! = \! 0.8$, which has a conspicuously small value of $C_{700}/C_{100}$.

The minimum subhalo mass motivated by CDM models is several orders of magnitude smaller than the smallest $M_{\rm min}$ used in this calculation, so it is worth considering the effect of smaller mass clumps on the power spectrum and its detectability.  Extending the mass function to smaller masses would boost the overall intensity of the emission, but estimates for the increase in intensity vary depending on the assumed structural properties of the subhalos and mass function.  For example, \citet{diemand_kuhlen_madau_07a} find the annihilation luminosity  is approximately constant per decade of subhalo mass, which results in a factor of $\sim 3$ greater total halo luminosity than the smooth halo alone, while the model presented by \citet{colafrancesco_profumo_ullio_06} implies a total boost of about 8 for their data.  Empirically, between subhalo masses of $10^{7}$ M$_{\odot}$ and 10 M$_{\odot}$ I find that the flux contribution roughly doubles with each decade of decreasing subhalo mass, which suggests that including emission from subhalos between $10^{-2}$ M$_{\odot}$ and 10 M$_{\odot}$ would increase the total substructure flux by a factor of $\sim 8$ over the M$_{\rm min} = 10$ M$_{\odot}$ substructure flux\footnote{The total contribution from subhalos with masses below $10^{-2}$ M$_{\odot}$ is found to be small by both \cite{colafrancesco_profumo_ullio_06} and \cite{diemand_kuhlen_madau_07a}.}.  A factor of a few boost in the overall intensity could improve the detection prospects for this more realistic CDM scenario as long as the additional flux does not overwhelm the anisotropy signal.
For the power spectrum, an extrapolation of the trends found here to $M_{\rm min} \! \ll \! 10$ M$_{\odot}$ implies that the power spectrum would continue to flatten at large multipoles and the overall amplitude would decrease.  

From Figure~\ref{fig:cl_ratio_compare} it is apparent that the smooth halo emission would be an overwhelming contaminant of the signal for the realizations with $M_{\rm min} \! = \! 10^{7}$ M$_{\odot}$ in the anti-biased distribution, but only a moderate ($N_{\rm b}/N_{\rm s} \! < \! 10$) or even negligible contaminant for all other scenarios.  The EGRB above 10 GeV as determined by \cite{strong_moskalenko_reimer_04} would exceed the dark matter emission from all realizations by at least a factor of $\sim 10$ assuming the adopted dark matter particle model.  The EGRB contribution to the measured power spectrum would in this case no longer be negligible relative to the substructure signal, but for many scenarios would be of a similar level.  Using spectral information to disentangle the EGRB or raising the energy threshold above 10 GeV to reduce the EGRB contamination could improve a measurement of the substructure power spectrum in this situation.

Fermi will primarily operate in sky-scanning mode, so the observing strategy is fixed, but the optimal strategy for extracting the signal from the data is worth considering.  With very limited information about the expected backgrounds this study has taken a naive approach, using as much of the data as possible outside of regions expected to be highly contaminated to maximize the number of signal photons collected, and restricting the data to energies above 10 GeV to further reduce contaminants.  However, this approach could certainly benefit from refinement once the properties of the diffuse emission are better known.  For example, the signal-to-noise variation across the map for different energy ranges will be determined by the foreground cleaning, and could then be used to select data to minimize the measurement uncertainties.
In addition, as suggested above, incorporating spectral information could strengthen this approach since the energy dependence of the signal from dark matter emission differs from that of expected astrophysical backgrounds.  Here I considered only the angular power spectrum of the continuum emission, but many WIMP candidates may also produce line emission by annihilating directly into $\gamma\gamma$ or $Z\gamma$ states.  Angular correlation of the line emission could help to strengthen an identification of the source as dark matter.

I have shown that for several scenarios Fermi could measure the angular power spectrum accurately enough to constrain the substructure population.  In general, the ability of Fermi to detect emission from dark matter is optimized for a particle with $m_{\chi} \lesssim 100$ GeV, based on Fermi's effective area and energy threshold.  Atmospheric Cherenkov Telescopes (ACTs) have higher energy thresholds but much larger effective areas, and hence would be more suitable for detecting gamma-rays from annihilation of more massive dark matter particles.  ACTs also generally have better angular resolution than Fermi, and could extend a measurement of the angular power spectrum to even higher multipoles. Unfortunately, ACTs must contend with an enormous flux of high energy cosmic-rays which can be difficult to distinguish from gamma-rays and present a formidable challenge to measuring a diffuse gamma-ray background with those experiments.  To adequately measure the angular power spectrum with ACTs, the contamination of the diffuse background would for most scenarios need to be reduced to the point that $N_{\rm b}/N_{\rm s}$ is less than $\sim 10$, a level beyond the reach of current experiments.  Even so, cross-correlation of the gamma-ray signal with data in other energy ranges should be pursued, both to boost the prospects of detecting a signal and to confirm or rule out the source of the signal as dark matter annihilation.

\section{Discussion}
\label{sec:discussion}

A confident identification of dark matter as the source of an observed signal is one of the greatest challenges for indirect detection.  The angular power spectrum of the diffuse gamma-ray background is an unique diagnostic of the properties of the source population, and as such could strengthen other approaches by providing complementary information.  

The angular power spectrum is determined solely by the angular distribution of the measured emission.  This study took advantage of this characteristic to explore the potential for constraining the small-scale dark matter distribution with a measurement of the angular power spectrum by Fermi.  Since expectations for the dark matter distribution on subgalactic scales depend on assumptions about the nature of dark matter and the adopted cosmology, predictions for the angular power spectrum of the emission are subject to considerable uncertainties.  
The models for the subhalo population considered here sample a few plausible scenarios, and while they certainly do not represent the full range of possibilities, they serve to illustrate that predictions for the small-scale distribution of dark matter could be tested by Fermi. 

The mass functions, density profiles, and radial distributions used in this work are motivated by simulations in a $\Lambda$CDM cosmology.  To approximate the subhalo population for other scenarios, such as SIDM or a suppression of power at small-scales in the primordial power spectrum, I employed a simple cut-off in the subhalo mass function below roughly the mass of a dwarf galaxy.  Although for these models the abundance of subhalos would be drastically reduced compared to the standard CDM case, a more sophisticated modification of the mass function would more accurately describe the subhalo population in these scenarios.  The sensitivity of the angular power spectrum to the assumed mass function should be studied in more detail as predictions from simulations which adopt alternate cosmologies are further refined.

My treatment did not account for the modification of the internal structure of the subhalos expected in SIDM models or from tidal stripping or stellar encounters.  For SIDM, the softening of the inner density profile may lead to a significant reduction in the total flux of a subhalo, but since since the angular size of the emission region of a subhalo is typically far smaller than the resolution limit of Fermi, I expect that its primary effect would be to reduce the overall intensity of the signal without altering the measured angular distribution of the emission significantly.  Similarly, dynamical processing of the subhalos is also likely to reduce the intensity of the emission slightly.  If subhalos are more efficiently disrupted in some regions of the Galaxy than others \cite{berezinsky_dokuchaev_eroshenko_07}, the measured angular power spectrum could vary between different patches of the sky.  

I considered two very different radial distributions in an attempt to bracket the most extreme cases, and found little difference in the angular power spectrum calculated for the entire sky in the relevant multipole range.  Although the anti-biased distribution does not show any large-scale dependence, for the unbiased distribution the angular power spectrum would likely differ somewhat across the sky due to the variation in subhalo number density.  
In addition, I assumed that the radial distribution of the subhalos is spherically symmetric about the Galactic Center, but asymmetric distributions are often found in simulations \citep[e.g.,][]{zentner_kravtsov_gnedin_etal_05}, and also could lead to a directional dependence of the angular power spectrum.  More detailed models of the subhalo spatial distribution incorporating the effects of tidal disruption \citep[e.g.,][]{berezinsky_dokuchaev_eroshenko_07} and correlations between halo concentration and radial distribution \citep[e.g.,][]{diemand_kuhlen_madau_07b} have been developed.  These models may be able to more accurately predict the directional dependence of the power spectrum. 

Uncertainties in the flux and spectrum of the gamma-ray emission from both dark matter and other source classes could make the overall amplitude of the substructure power spectrum difficult to determine, so initially the shape of the power spectrum may be a better indicator of the abundance of substructure than the measured amplitude.  I note, however, that the energy dependence and absolute amplitude of the power spectrum could provide useful information, particularly once emission from other source classes is sufficiently constrained.

This work reaffirms the findings of \cite{kuhlen_diemand_madau_08} that the spatial distribution and intensity of the gamma-ray emission from Galactic dark matter annihilation are strongly affected by the radial distribution of the subhalo population.  An anti-biased subhalo distribution results in considerably less total emission than an unbiased distribution due to the larger typical distances of the subhalos, and in the case of a small minimum subhalo mass as generically predicted for CDM, the observed emission is almost entirely diffuse and devoid of large-scale features.  

In addition to impacting the detectability of individual subhalos \cite{kuhlen_diemand_madau_08}, an anti-biased radial distribution may alter the prospects for detecting the annihilation signal by several other methods.
The prominence of the dipole from the Galactic dark matter halo  would be diminished relative to the unbiased distribution.   Predictions for scenarios in which the substructure mass function extends to very small masses would be most strongly impacted.  The additional flux from the outer regions of the halo would affect the angular dependence of the emission by increasing the mean observed intensity from dark matter annihilation in all directions. 
A related consequence concerns constraints on extragalactic dark matter emission derived from the detectability of the Galactic Center.  \citet{ando_05} presented an argument using self-similarity of halo structure to constrain the fraction of the EGRB due to dark matter annihilation using the signal from the Galactic Center.  However, if a substantial fraction of the luminosity of a galaxy due to dark matter annihilation originates from the outer regions of the halo, as would be the case for an anti-biased subhalo distribution, the flux from the Galactic Center as measured from our position would provide only a weak constraint.
A better understanding of these issues will be needed to clarify the interpretation of a Fermi detection (or non-detection) of predicted dark matter signals.

\section{Summary}
\label{sec:summary}

With the Fermi era just around the corner, this study and many others have focused on the exciting possibility that dark matter may soon be robustly detected for the first time via gamma-ray emission.
In this work I considered the angular power spectrum of diffuse gamma-ray emission as a means of constraining the abundance and properties of dark matter substructure.  Using models motivated by numerical simulations to describe the Galactic dark matter distribution, I calculated the angular power spectrum from dark matter annihilation in substructure for several scenarios and evaluated the prospects for measuring this signal and distinguishing between dark matter models with observations by Fermi.  

I showed that the presence of Galactic dark matter substructure  induces distinct features in the angular power spectrum of diffuse gamma-ray emission.  A cut-off in the substructure mass function below roughly the mass of known dwarf galaxies results in declining power ($C_{\ell}$) in the multipole range $100 \lesssim \ell \lesssim 1000$.  In contrast, if the substructure mass function extends down to the minimum masses predicted for CDM halos, the angular power spectrum coefficients $C_{\ell}$ of the diffuse emission will be almost constant over this multipole range.  

For several models, I found that a measurement of the angular power spectrum by Fermi will be able to differentiate between a halo populated by a multitude of subhalos and one in which substructure is scarce, as long as the contamination of the dark matter signal by the EGRB is not overwhelming.  The detectability of the signal is determined largely by the subhalo abundance and radial distribution, which can change the mean intensity of the emission from dark matter annihilation by several orders of magnitude.

The gamma-ray emission produced by an anti-biased subhalo distribution differs greatly in intensity and spatial distribution from that produced by an unbiased distribution.  An anti-biased distribution generates emission of similar intensity in all directions, with fluctuations confined to small angular scales, and does not enhance the large-scale dipole feature from the smooth halo component as an unbiased distribution does.  The detectability of dark matter annihilation by a variety of methods may be altered relative to the case of an unbiased distribution if the subhalo distribution is anti-biased.

\ack
I am greatly indebted to Angela Olinto for insights and guidance over the course of this work, and to Andrey Kravtsov for valuable advice and suggestions.  I also thank Maximo Ave, John Beacom, Brian Fields, Stephan Meyer, Vasiliki Pavlidou, Luis Reyes, Simon Swordy, and Carlos Wagner for helpful discussions.  This work was supported by NSF grant PHY-0457069 and by the Kavli Institute for Cosmological Physics at the University of Chicago through grants NSF PHY-0114422 and NSF PHY-0551142 and an endowment from the Kavli Foundation and its founder Fred Kavli.

\appendix

\section*{Appendix. Comparison of the annihilation flux from the NFW and Einasto density profiles}
\label{appendix}
\setcounter{section}{1}

The flux from dark matter annihilation from a subhalo at a distance $d$ is
\begin{equation}
\label{eq:dmflux}
F=\frac{K}{4 \pi d^{2}}\int {\rm d}V\, \rho^{2}(r),
\end{equation}
where the integration is over the subhalo volume $V$ and $\rho(r)$ is the density profile of the subhalo.
Using the Einasto density profile [equation (\ref{eq:einden})] and integrating the volume enclosed within $r_{200}$, the flux from a subhalo is
\begin{equation}
\label{eq:einflux}
F_{\rm \scriptscriptstyle Ein}= \frac{K\, \rho_{s}^{2}\, r_{s}^{3}\, e^{^{4/\alpha}}}{4d^{2}} \left(\frac{\alpha}{4}\right)^{^{\frac{3}{\alpha}-1}}\!\! \gamma\!\left[{\textstyle \frac{3}{\alpha}, \frac{4}{\alpha} (c_{200})^{^{\alpha}}}\right]
\end{equation} 
where $\gamma[a,x] \! = \! \int_{0}^{x}{\rm d}t\,\, t^{a-1}e^{-t}$ is the lower incomplete gamma function.
For the NFW profile [equation (\ref{eq:nfwden})], the corresponding flux is
\begin{equation}
\label{eq:nfwflux}
F_{\rm \scriptscriptstyle NFW}= \frac{K\, \rho_{s}^{2}\, r_{s}^{3}\,}{3d^{2}} \left(1-\frac{1}{(1+c_{200})^{3}}\right).
\end{equation}

\begin{figure}[ht]
	\centering
	\includegraphics[width=0.9\textwidth]{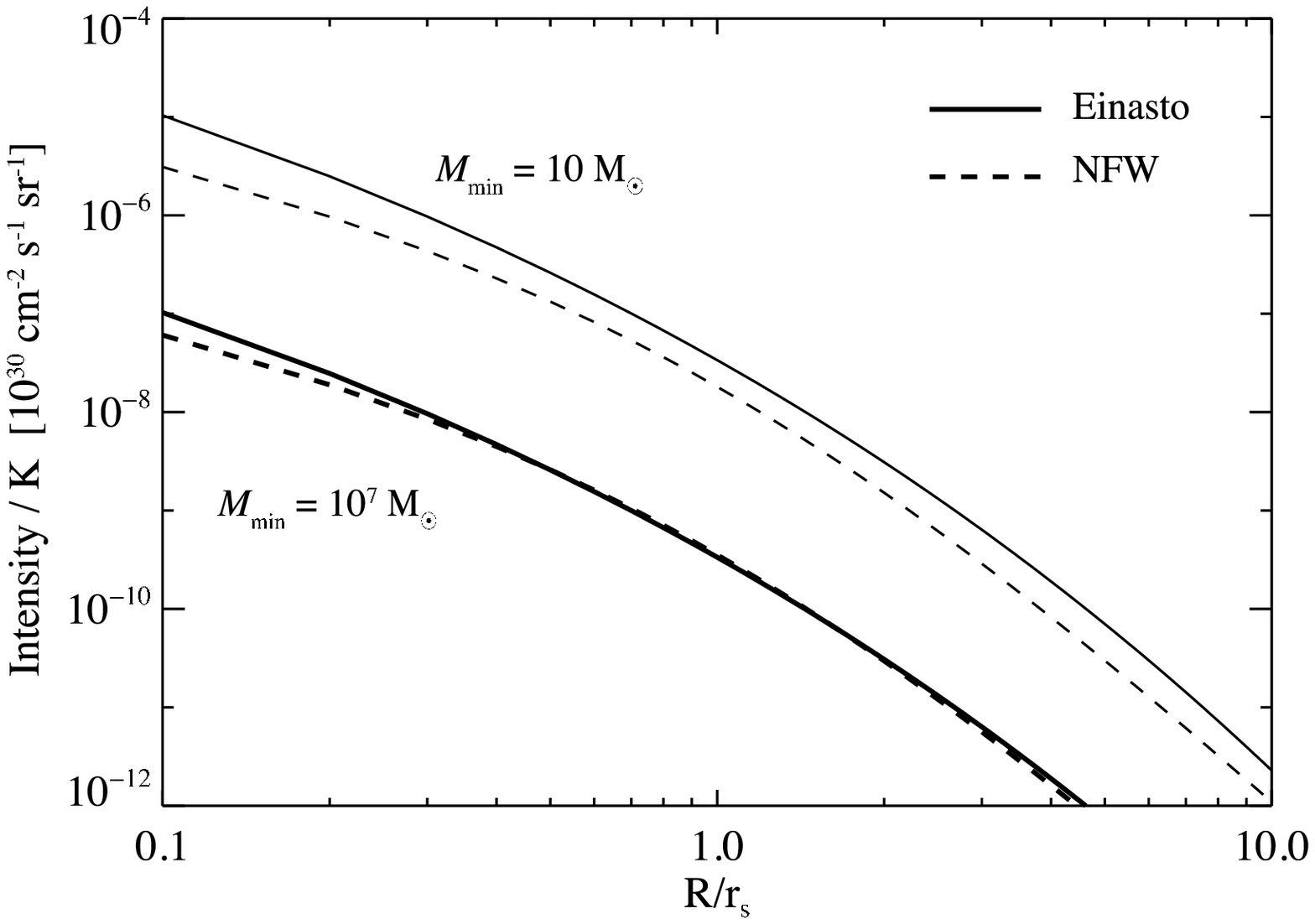}
	\caption{Dependence of the gamma-ray intensity on projected radius from the center of the subhalo for the Einasto ({\it solid lines}) and NFW ({\it dashed lines}) density profiles.  The thick (thin) lines are for a $10^{7}$ (10) M$_{\odot}$ subhalo.
	\label{fig:ein_nfw_intens_compare}}
\end{figure}

Using the mass-concentration relation given in equation (\ref{eq:c200_m200_rel}), for the same subhalo mass the Einasto profile results in a slightly larger total flux than the NFW profile (larger by a factor of~$\sim\! 1.5$ for a $10^{7}$ M$_{\odot}$ subhalo,~$\sim\! 3$ for a 10 M$_{\odot}$ subhalo).  The intensity as a function of projected radius in units of the scale radius $r_{\rm s}$ is shown in Figure~\ref{fig:ein_nfw_intens_compare}.  The increased intensity from the Einasto profile is primarily in the inner regions.  For both profiles, the majority of the subhalo flux originates from within a projected radius of $R \! = \! 1 r_{\rm s}$ ($\sim \! 92$\% of the flux for the NFW profile,~$\sim\! 95$\% for the Einasto profile).

\bibliography{anisotropies}   

\end{document}